\documentclass[prd,superscriptaddress,twocolumn,floatfix,nofootinbib]{revtex4}
\usepackage{epsfig} 
\usepackage{color}
\usepackage{bm}

\def\ds{\displaystyle}
\def\Sun{\odot}
\def\sun{\hbox{$\odot$}}
\def\hmpcinv{h\,{\rm Mpc}^{-1}}
\def\hmpc{h^{-1}\,{\rm Mpc}}
\newcommand{\fnl}{f_{\rm NL}}
\newcommand{\zmax}{z_{\rm max}}

\def\eg{{\it e.g.~}}
\def\etal{{\it et al.~}}

\def\ben{\begin{enumerate}}
\def\een{\end{enumerate}}
\def\bi{\begin{itemize}}
\def\ei{\end{itemize}}
\def\be{\begin{equation}}
\def\ee{\end{equation}}
\def\bea{\begin{eqnarray}}
\def\eea{\end{eqnarray}}
\newcommand{\comment}[1]{}
\newcommand{\apjl}{\apj\ Lett.}
\newcommand{\apjs}{\apj\ Supp.\ Ser.}
\newcommand{\mnras}{Mon.\ Not.\ R.\ Astron.\ Soc.} 

\begin{document} 
\title{The imprints of primordial non-gaussianities on large-scale structure:
scale dependent bias and abundance of virialized objects}
\author{Neal Dalal}
\author{Olivier Dor\'{e}}
\affiliation{
Canadian Institute for Theoretical Astrophysics, 60 St.\ George St, 
University of Toronto, Toronto, ON, Canada M5S3H8} 
\author{Dragan Huterer}
\affiliation{
Kavli Institute for Cosmological Physics and Department of Astronomy 
and Astrophysics, University of Chicago, Chicago, IL 60637}
\affiliation{
Department of Physics, University of Michigan, 450 Church St, 
Ann Arbor, MI 48109}
\author{Alexander Shirokov}
\affiliation{
Canadian  Institute for Theoretical Astrophysics, 60 St.\ George St, 
University of Toronto, Toronto, ON, Canada M5S3H8} 

\received{\today}
%
\begin{abstract}
We study the effect of primordial nongaussianity on large-scale
structure, focusing upon the most massive virialized objects.  Using
analytic arguments and N-body simulations, we calculate the mass
function and clustering of dark matter halos across a range of
redshifts and levels of nongaussianity.  We propose a simple fitting
function for the mass function valid across the entire range of our
simulations.  We find pronounced effects of nongaussianity on the
clustering of dark matter halos, leading to strongly scale-dependent
bias.  This suggests that the large-scale clustering of rare objects
may provide a sensitive probe of primordial nongaussianity.  We very
roughly estimate that upcoming surveys can constrain nongaussianity at
the level $|\fnl|\lesssim 10$, competitive with forecasted constraints
from the microwave background.
\end{abstract}
\maketitle
\section{Introduction}

One of the fundamental predictions of standard (single-field, slow-roll)
inflationary cosmology is that
the density fluctuations in the early universe that seeded large-scale
structure formation were nearly gaussian random (e.g.\
\cite{maldacena,Acquaviva:2002ud,Creminelli:2003iq,Lyth_Rodriguez,Seery_Lidsey}).
Constraining or detecting non-gaussianity (NG) is therefore an important and
basic test of the cosmological model.  To the extent that it can be measured,
gaussianity has so far been confirmed; the tightest existing
constraints have been obtained from observations of the cosmic
microwave background \cite{wmap3,Creminelli_wmap}. Recently, several
inflationary models have been proposed which predict a potentially
observable level of nongaussianity, see \eg 
\cite{ArkaniHamed:2003uz,Bartolo:2003jx,Lyth:2005du,Rigopoulos:2005ae,
Allen:2005ye,Chen_DBI,Barnaby:2006cq,Barnaby:2006km,Barnaby:2007yb,
Sasaki:2006kq,Chen:2006nt,Chen_Easther_Lim,Battefeld_Easther,
Assadullahi:2007uw,Battefeld:2007en,Shandera} 
and \cite{Bartolo:2004if} for a review.
Improved limits on NG would rule out some of these models; conversely, a robust
detection of primordial nongaussianity would dramatically overturn standard
inflationary cosmology and provide invaluable information about the nature of
physical processes in the early universe. In this regard, there has been a
resurgence in studying increasingly more sophisticated methods and algorithms
to  constrain (or, if we are lucky, detect) nongaussianity
\cite{Babich:2005en,Babich_shape,Creminelli_estimators,Smith_Zaldarriaga,Fergusson_Shellard}.

Nongaussianity manifests itself not only in the cosmic microwave background
\cite{Falk_Ran_Sre,Luo_Schramm,Gangui_etal,Wang_Kam}, but also in the late-time
evolution of large-scale structure.  For example, detailed measurements of
higher order correlations like the bispectrum or trispectrum of galaxy
clustering could provide a handle on primordial nongaussianity
\cite{Verde:2000vr,Scoccimarro:2003wn,Sefusatti:2007ih}.  The abundance of
galaxy clusters, the largest virialized objects in the universe, has also long
been recognized as a sensitive probe of primordial NG
\cite{Lucchin:1987yv,Robinson:1999se,Benson:2001hc,Matarrese:2000iz,verde01,Scoccimarro:2003wn,Komatsu:2003fd}.  Because clusters are rare objects which form from the
largest fluctuations on the tails of the density probability distribution,
their abundance is keenly sensitive to changes in the shape of the PDF such as
those caused by nongaussianity.  Large statistical samples of massive clusters
have already been compiled from wide-area optical imaging and spectroscopic
surveys such as the Sloan Digital Sky Survey \cite{maxBCG,Koester:2007bg}, the
Two-Degree Survey \cite{Eke:2004ve}, and from the Red Sequence Survey
\cite{Yee:2007if} and from X-ray surveys using the Chandra and XMM-Newton
observatories \cite{Willis:2005ag,Valtchanov:2003it}.  Future missions, such as
the Dark Energy Survey, Supernova/Acceleration Probe and Large Synoptic Survey
Telescope, will detect and study tens of thousands of clusters, revolutionizing
our understanding of cluster physics as well as providing important constraints
on cosmology \cite{Haiman_Mohr_Holder,Majumdar_Mohr,Wang_Haiman,
Battye_Weller,Lima_Hu_05, Marian_Bernstein,Takada_Bridle}.

To exploit the potential of these upcoming surveys as probes of
primordial nongaussianity, it is important to calibrate the effects of
NG on the abundance and clustering of virialized objects.  While no
previous work has attempted to quantify the effects of NG on halo
clustering, several groups over the past decade have constructed
fitting formulae for the halo mass function
\cite{Robinson_Baker,Robinson_Gawiser_Silk,MVJ}.  All of this work,
however, was analytic and relied on the validity of the Press-Schechter
\cite{press-schechter} formalism, plus various further approximations.
The resulting analytic estimates are, in general, rather cumbersome to
compute and have questionable accuracy.  As discussed below, the
Press-Schechter model provides only a qualitative description of halo
abundance, and fails to reproduce the halo mass function to within an
order of magnitude over the mass and redshift range accessible to
current and future cluster surveys.  Therefore, analytic models for NG
cluster abundance based on the Press-Schechter ansatz may not be
sufficiently accurate.  Given the high-quality data soon to be
available, a much more precise calculation of cluster statistics will
be required.  Quite recently, two groups have attempted to
quantify the mass function of clusters in NG models using N-body
simulations \cite{Kang,Grossi}, reaching contradictory conclusions.

In this paper, we use analytic arguments and
numerical simulations to estimate the effect of
NG on the abundance and clustering of virialized objects.
Because N-body simulations can be expensive and there is a wide NG parameter
space, we also strive to make our results useful to a cosmologist who
is not necessarily equipped with the machinery or patience to run simulations or
evaluate difficult analytic expressions. To this end, we provide a simple,
physically motivated fitting formula for the halo mass function and
halo bias, which we calibrate to our N-body simulations.

Our main results are that the mass function and correlation function
of massive halos can be significantly modified by primordial
nongaussianity.  We find a somewhat weaker effect of NG on the mass
function than previous analytic estimates.  We also show analytically
and numerically that NG strongly affects the clustering of rare
objects on large scales, implying that measurements of the large-scale
power spectrum can place stringent bounds on NG.  

The plan of the paper is as follows.  In Section \ref{anal} we derive
analytic expressions for the abundance and clustering of rare peaks.
In Section \ref{sec:method} we describe our N-body simulations, followed
in Section \ref{sec:mf} by a discussion of our measured halo mass
function, and our fitting formula for the mass function.  In Section
\ref{sec:pk} we present measurements of halo clustering within our
simulations, and in Section \ref{sec:cosmo} we discuss cosmological
implications of our findings.  

\section{Analytic estimates} \label{anal}

In this section, we derive analytic expressions for the abundance and
clustering of dark matter halos.  As mentioned above,
such analytic approaches provide a useful qualitative framework for
understanding gravitational collapse, however they
cannot be used to describe quantitatively either the mass function or
the clustering amplitude of collapsed objects.  The expressions
derived here are meant solely to motivate the more precise fitting
formulae described in subsequent sections.

We will focus on local NG of the form \cite{MVJ,komatsu,maldacena}
\begin{equation}
\Phi_{\rm NG}({\bm x})=\phi({\bm x}) + \fnl 
(\phi^2({\bm x})-\langle\phi^2\rangle).
\end{equation}
In our notation, $\Phi=-\Psi$, where $\Psi$ is the usual Newtonian
potential. On subhorizon scales, this choice of Newtonian gauge is
valid, and the potentials $\Phi$ and $\Psi$ satisfy the Poisson
equation relating them to the overdensity $\delta$.  On superhorizon
scales, the Bardeen potential $\Phi$ and overdensity $\delta$ are
proportional, and not related by a Poisson
equation, so our analysis will be valid only on subhorizon scales.
With this choice of convention, positive $\fnl$ corresponds to
positive skewness of the density probability distribution, and hence
an increased number of massive objects.

For simplicity, we neglect the effect of the CDM transfer functions,
which modify the shape of the $\Phi$ power spectrum after
nongaussianity is generated.  Then the
probability distribution for $\Phi_{\rm NG}$ is easy to write
down, however the probability distribution for the density 
$\delta_{\rm NG}$ cannot be expressed analytically.  Nevertheless, we
can make progress by assuming that the NG correction is small, and by
focusing only on high peaks of the density.  
The Laplacian of $\Phi_{\rm NG}$ is 
\begin{equation}
\nabla^2 \Phi_{\rm NG} = \nabla^2\phi + 2\fnl [\phi\nabla^2\phi+|\nabla\phi|^2].
\label{grad2phi}
\end{equation}
Because $\phi$, $\nabla\phi$, and $\nabla^2\phi$ are all Gaussian
fields whose statistics are fully specified by their power
spectra, then Eqn.~(\ref{grad2phi}) above, relating
$\delta_{\rm NG}=-(3\Omega_m/2ar_H^2)\nabla^2\Phi_{\rm NG}$ to the
Gaussian fields, allows us to determine fully the statistics of the
nongaussian density $\delta_{\rm NG}$.  For example, the skewness of
$\delta_{\rm NG}$ becomes, to lowest order in $\fnl$,
\begin{equation}
S_3=\frac{\langle\delta_{\rm NG}^3\rangle}{\langle\delta_{\rm NG}^2\rangle^2}
=6\fnl\frac{\langle\phi\delta\rangle}{\sigma_\delta^2}.
\end{equation}

On the average, the two terms $\phi\nabla^2\phi$ and $|\nabla\phi|^2$
in Eqn.~(\ref{grad2phi}) are of the same order; the fact that they
have equal but opposite expectation value is why 
$\langle\delta_{\rm NG}\rangle=\langle\delta\rangle=0$.  
However we are mainly interested in high peaks, where
$\delta\propto-\nabla^2\phi$ is large.  Because $|\nabla\phi|^2$ is
uncorrelated with $\nabla^2\phi$, and because at the peak of $\phi$
its derivative vanishes, we assume that $|\nabla\phi|^2$ may be neglected
compared to $\phi\nabla^2\phi$ in the vicinity of rare, high peaks.
Then applying the Poisson equation near the peak gives 
$\delta_{\rm NG}\approx \delta [1+2\fnl\phi]$.  This expression
applies for the primordial density and potential fields at early
times.  At late times, $\delta_{\rm NG}$ subsequently grows according
to the linear growth factor $D(a)$, while the potential decays like
$g(a)\propto D(a)/a$. Therefore, rewriting this expression in terms of
the late-time fields, we find
\begin{equation}
\delta_{\rm NG}\approx \delta [1+2\fnl\phi/g(a)].
\label{dNG}
\end{equation}
We see that the peak height is enhanced by a factor proportional to
the primordial potential $\phi_p=\phi/g(a)$, rather than the evolved
potential.\footnote{An earlier version of this paper neglected to
  distinguish between the primordial and late-time potential, and
  hence omitted the $g(a)$ factor.  We are grateful to N.\ Afshordi
  for pointing this out to us.}

Equation~(\ref{dNG}) will be the basis for the rest of our discussion.
We emphasize that this is only valid in the vicinity of peaks, and so 
we focus on peaks for the remainder of this
discussion.  Because the fields $\delta$
and $\phi$ are Gaussian distributed, we can immediately derive
properties of the distribution of $\delta_{\rm NG}$.  
For example, consider the mean shift in peak height for a peak
of Gaussian density $\delta$: 
\begin{eqnarray}
\langle\delta_{\rm NG}|\delta\rangle
&=&\delta\,(1+2\fnl\langle\phi_p|\delta\rangle)\nonumber\\
&=&\delta\,\left(1+2\fnl\frac{\langle\phi\delta\rangle}{g\sigma_\delta^2}
\delta\right)\;.
\end{eqnarray}
If the peak height $\delta$ and background potential $\phi$ were
uncorrelated, then there would be no systematic shift in peak height,
and hence no change in the abundance of massive halos.  However,
$\delta$ and $\phi$ are correlated, implying that rare peaks are
systematically raised or lowered, depending upon the sign of $\fnl$.
Therefore, we expect changes in the mass function and the correlation
function.

In the appendix, we derive expressions for the abundance and
clustering of regions above a given threshold, which then give the
clustering and mass function of halos in the Press-Schechter model.
However, we can derive the form of the halo correlation function using
a very simple argument.  The halo correlation function is usually
parameterized in terms of the halo bias $b$, which is the rate of
change of the halo abundance as the background density is varied.
Writing the matter overdensity as $\delta$ and the halo overdensity as
$\delta_h$, we can define the halo bias as
\begin{equation}
\delta_h = b\,\delta.
\end{equation}
It is normally assumed that $b\to$ const on large scales, but we will
not make this assumption here.  Consider a long-wavelength mode,
providing a background density perturbation $\delta$ and corresponding
potential fluctuation $\phi$.  In the absence on nongaussianity, this
perturbation raises subthreshold peaks above threshold, and thereby
enhances the abundance of super-threshold peaks by $b_L \delta$, where
$b_L$ is the usual (Gaussian) Lagrangian bias.  For nonzero $\fnl$,
the long-wavelength mode also enhances the peak height by
$2\fnl\phi_p\delta_{\rm pk}$, and we will focus on peaks near
threshold, such that $\delta_{\rm pk}\simeq\delta_c$.  This 
provides an additional enhancement factor, giving a total 
\begin{equation}
\delta_h = b_L (\delta + 2\fnl\phi_p\delta_c).
\end{equation}
In Fourier space, the potential and density modes are related by
$\phi = (3\Omega_m/2ar_H^2k^2)\delta$, and so we see that the
nongaussian bias acquires a correction
\begin{equation}
\Delta b(k)=2 b_L \fnl\delta_c \frac{3\Omega_m}{2ag(a)\,r_H^2k^2}\;,
\end{equation}
where again $b_L$ refers to the usual Lagrangian bias for halos of
this mass with Gaussian fluctuations.  The total Lagrangian bias is
then $b_L(k)=b_L + \Delta b(k)$.

Since we have been working with the clustering of peaks in the initial
density distribution, the
above expression for the bias applies only to the early-time,
Lagrangian bias.  Translating these results to
late-time, Eulerian bias is straightforward, however. The bias of
Eulerian halos is simply $b = 1+b_L$ : the excess of halos in some
Eulerian volume with overdensity $\delta$ is $b\delta = b_L\delta +
\delta$.  The first term corresponds to the excess of peaks in the
initial Lagrangian volume, which are advected into the Eulerian
volume.  The second term arises because an Eulerian volume with
overdensity $\delta$ has $\delta$ times more mass than an average
volume, and therefore $\delta$ times more peaks.  

In summary, local NG generates a scale-dependent correction to the
bias of galaxies and halos, of the form
\begin{equation}
\Delta b(k)=2(b-1)\fnl\delta_c \frac{3\Omega_m}{2a\,g(a) r_H^2k^2}
\label{bias_eul}
\end{equation}
where $b$ here now refers to the Eulerian bias of the tracer
population.  In subsequent sections, we show that this simple
expression, despite the underlying assumptions and 
approximations in its derivation,
matches surprisingly well the halo clustering measured in our
numerical simulations.

\section{Numerical simulations}\label{sec:method}

\begin{figure}[!t]
\epsfig{file=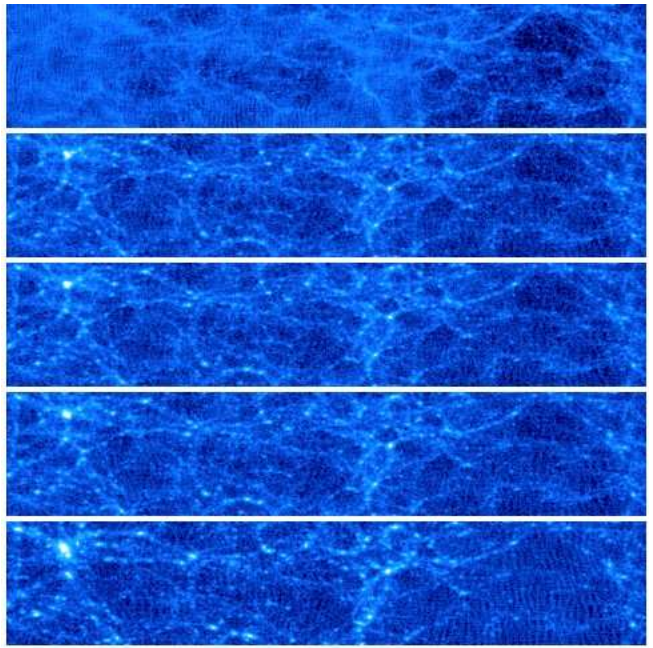,width=0.48\textwidth}
\caption{Slice through simulation outputs at $z=$ $0$ generated with the
  same Fourier phases but with $\fnl=$$-5000,$ $-500,$ $0,$ $+500,$ $+5000$
  respectively from top to bottom. Each slice is 375 $\hmpc$ wide, and 80 $\hmpc$ high
  and deep. We can easily match by eye much of the large scale
  structure; for example, an overdense region sits on the left, while
  an underdense region (void) falls on the right, in all panels.
  Note that for positive $\fnl$, overdense regions are more evolved
  and produce more clusters than their Gaussian counterparts, while
  underdense regions are less evolved (\eg grid lines are still
  visible).  For negative $\fnl$, underdense regions are more evolved,
  producing deeper voids, while overdense regions are less evolved, as
  illustrated by the grid lines apparent in the left of the top panel.}\label{fig:slice_sims}
\end{figure}

We numerically simulate the growth of structure in nongaussian
cosmologies using the adaptive P$^3$M parallel N-body code 
{\tt GRACOS}\footnote{http://www.gracos.org} 
\cite{gracos,alexthesis}.  Non-gaussian initial conditions were
generated using 
the following procedure.  First, we generated a Gaussian random
potential field $\phi({\bf x})$ using a power-law power spectrum with
a scalar (density) index $n_s=0.96$, and normalized so that
$\sigma_8=0.76$ \cite{wmap3} when multiplied by the matter transfer
function.  Following Refs.\ \cite{MVJ,komatsu,maldacena}, we then
computed the nongaussian potential $\Phi$ by adding a quadratic
correction in configuration space, 
\begin{equation}
\Phi({\bm x}) = \phi({\bm x}) +\fnl (\phi^2-\langle\phi^2\rangle).
\label{eq:fnl_def}
\end{equation}
We then multiplied $\Phi$ by matter transfer functions in Fourier
space for $\Omega_m=0.24$, $\Omega_\Lambda=0.76$, and computed
particle displacements and velocities using the Zeldovich
approximation \cite{padmanabhan}.

One immediate drawback to this approach is that, due to the strong
Fourier mode coupling generated by the $\fnl$ term, our results may be
affected by the absence of modes below the fundamental frequency or
above the Nyquist frequency of our simulation volume.  All N-body
simulations can cover only a finite dynamic range, and therefore have
zero power outside of their $k$-space volumes.  For Gaussian
simulations, this is believed not to be a serious defect, because mode
coupling is unimportant on linear scales, and on nonlinear scales, the
mode coupling generally transfers power to small scales.  In our case,
however, the $\fnl$ term couples all the modes sampled in our
simulation to all the modes absent in our simulation.  We have
performed rudimentary estimates of the magnitude of this effect, by
running simulations in which we high-pass or low-pass filter the
$\fnl$ correction, and do not observe significant changes in the
overall behavior.  Strictly speaking, however, it
must be borne in mind that our results apply only for power spectra
that are non-vanishing only over the finite range covered by our
simulation volume.  

We have performed several simulations using both Gaussian and
nongaussian initial conditions.  For each
Gaussian realization, we construct non-Gaussian realizations using the
same Fourier phases, with various $\fnl$, \eg $\fnl=\pm500$, $\pm50$,
and $\pm5$.  We ran simulations from a starting 
expansion factor $a=0.02$ until the present time, $a=1$, using $512^3$
particles in a box of sidelength $L=800 h^{-1}$ Mpc.  For these parameters,
each particle has a mass $m_p=2.52\times 10^{11} h^{-1}M_\odot$, so that
clusters with masses exceeding $M>10^{14} h^{-1}M_\odot$ are resolved with
$N\gtrsim 400$ particles.  Since we are interested mainly in the masses and
positions of cluster-sized halos, and not their internal structure, we have not
used high force resolution: we employ a Plummer softening length $l$ of 0.2
times the mean interparticle spacing.  We have checked that using higher force
resolution ($l$ half as large) does not appreciably change the mass function.
All simulations were performed at the Sunnyvale cluster at CITA; depending upon
the value of $\fnl$, the simulations completed in 2-3 hours each on
typically 8-10 nodes.  As a consistency check, we have also run a
small number of $1024^3$ particle simulations with the same particle
mass and force softening as above, but with twice the box size.  These
larger runs typically completed in 18-20 hours on 64 nodes.  In
Figure~\ref{fig:slice_sims}, we plot slices through our simulation volume at
redshift $z=0$, and the effects of varying $\fnl$ are readily apparent.  Large
positive $\fnl$ accelerates the evolution of overdense regions and retards the
evolution of underdense regions, while large negative $\fnl$ has precisely the
opposite effect.

\section{The halo mass function} \label{sec:mf}

\begin{figure}[!t]
\epsfig{file=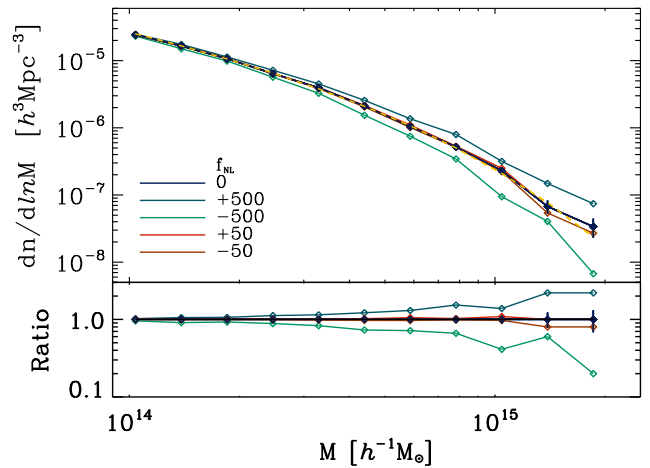,width=0.37\textwidth,angle=+90}
\caption{Mass functions measured from simulations with various $\fnl$
  and identical phases (3 sets of initial conditions were used for
  each $\fnl$).
  The top panel shows the mass function as well as the Gaussian
  fitting formula (dashed yellow line) from \citet{warren}. 
  The bottom panel shows the ratio between the measured $\fnl=0$
  Gaussian mass functions and the respective non-Gaussian ones. 
\label{fig:mf_warren}}
\end{figure}

We constructed late-time halo catalogues at redshifts $z=1$, 0.5, and
0 using the friends-of-friends group finder \cite{davis}, with linking
length $b=0.2$.  For Gaussian simulations, the halo mass function
constructed this way has been extensively calibrated
\cite{jenkins,warren}.  Resulting mass functions are plotted in 
Figure~\ref{fig:mf_warren}.  

\subsection{A new fitting formula}\label{sec:fit_formula}

Having measured the halo mass function, we next would like to construct a
fitting function along the lines of those used for Gaussian simulations
\cite{jenkins,warren}.  As mentioned above, previous techniques for estimating
the nongaussian mass function have been based upon the Press-Schechter
\cite{press-schechter} ansatz.  Given that the Press-Schechter mass function
fails to match the halo mass function to within an order of magnitude over the
mass and redshift ranges of interest to us \cite{warren}, and given the lack of
any physical basis to the Press-Schechter ansatz \cite{bcek, bm96}, we have
instead adopted an alternative approach which we describe next.

We start by noting that the halo mass function $dn/dM$ has been
precisely calibrated for Gaussian cosmologies.  Consider a Gaussian
realization of the density field, which at late times evolves to
produce halos with mass function $dn/dM_0$. As we slowly vary
$\fnl$ away from zero, the structures forming at late times also
slowly vary (c.f.\ Figure \ref{fig:slice_sims}), 
producing a different mass spectrum $dn/dM_f$. 
If we vary $\fnl$ slowly enough, we can track the change in mass and
position for individual halos: i.e., for each halo of mass $M_0$ for
$\fnl=0$, we can uniquely identify a corresponding halo of mass $M_f$
for $\fnl\ne 0$, as long as $|\fnl|$ is sufficiently small.  Since we
know precisely the number of halos as a function of $M_0$, if we
can determine the mapping $M_0\rightarrow M_f$, we will then have an
estimate of the non-Gaussian mass function $dn/dM_f$ via

\begin{equation}
\frac{dn}{dM_f} = \int dM_0 \frac{dn}{dM_0} \frac{dP}{dM_f}(M_0),
\label{eq:mf_conv}
\end{equation}
where $dP/dM_f(M_0)$ is the probability distribution that a Gaussian
halo of mass $M_0$ maps to a non-Gaussian halo of mass $M_f$.  Note
that the probability distribution function $dP/dM_f$ need not
integrate to unity, $\int dM_f\,dP/dM_f\ne 1$ in general, since the
total number of halos is not conserved: halos can merge or split as
$\fnl$ is varied.

The next step is to determine the probability distribution
$dP/dM_f(M_0)$, by matching halos between Gaussian and non-Gaussian
simulations.  We match halos by requiring that matching pairs have
significantly overlapping Lagrangian volumes; i.e.\ by requiring that
halos have many particles in common, where particles are labeled by
their Lagrangian coordinates in the initial conditions.  For each halo
$M_f$ in a non-Gaussian run, we loop over the halo's particles and
identify which Gaussian halos own those particles in the run with
$\fnl=0$.  The Gaussian halo owning the largest fraction (exceeding
1/3) of the particles is then identified as the match for non-Gaussian
halo $M_f$.  Each Gaussian halo $M_0$ can have one, several, or zero
matching non-Gaussian halos, depending on $\fnl$.  By stacking
Gaussian halos of similar mass $M_0$, we can determine
$dP/dM_f(M_0)$. 

\begin{figure}[!t]
\centerline{
\epsfig{file=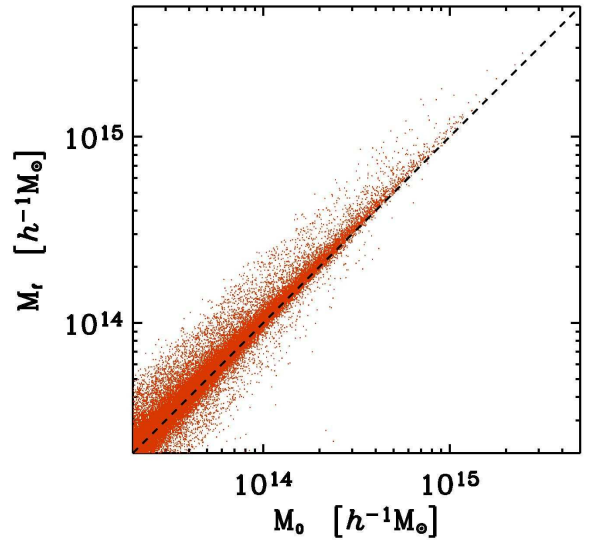,width=0.45\textwidth,angle=+90}
}
\caption{Distribution of $M_f$ as a function of $M_0$ for one 
  $\fnl=+500$ simulation. The average shift towards higher masses
  is clearly visible.}
\label{fig:scatter_momf}
\end{figure}

\begin{figure*}[!t]
\epsfig{file=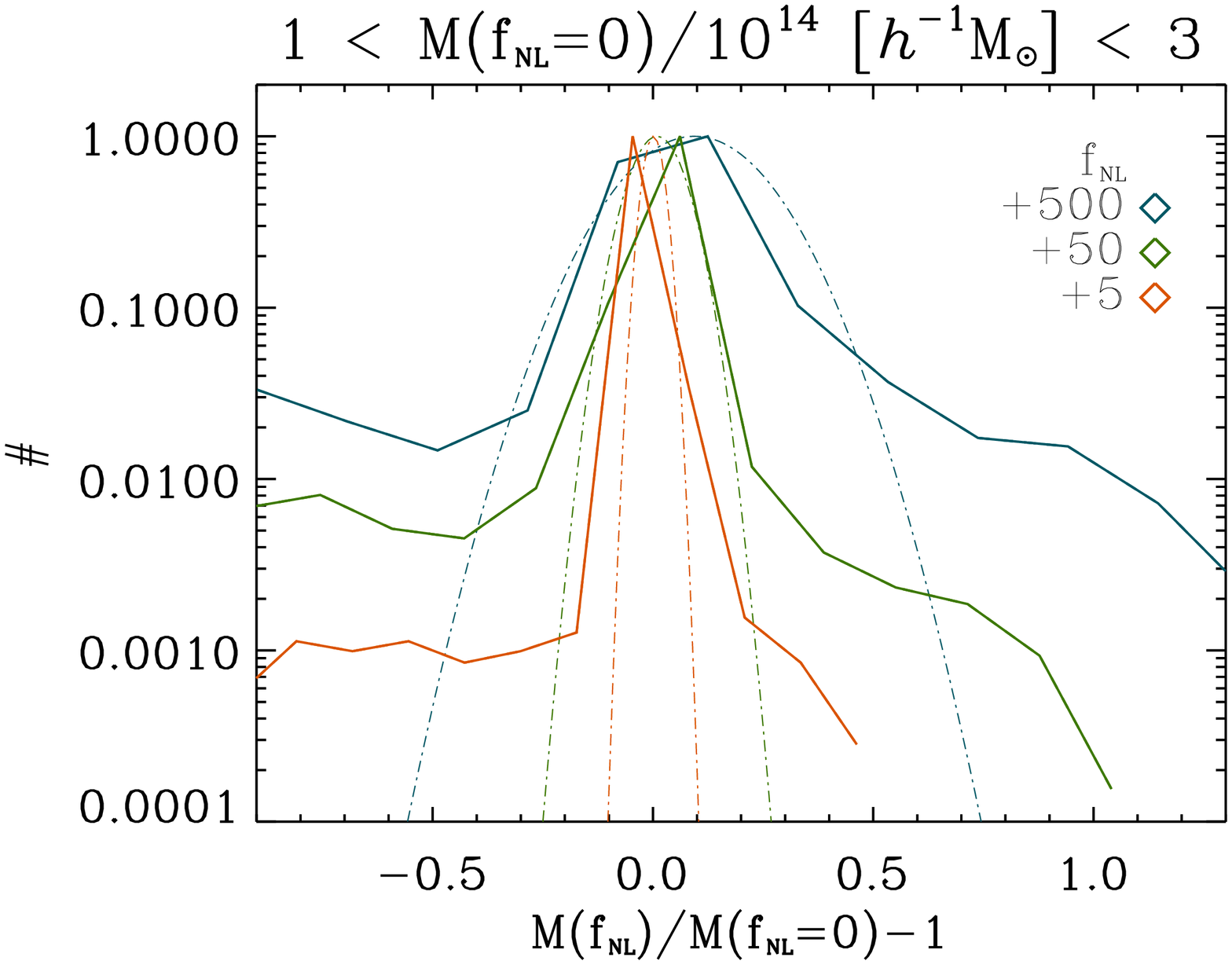,width=0.36\textwidth,angle=+90}
\epsfig{file=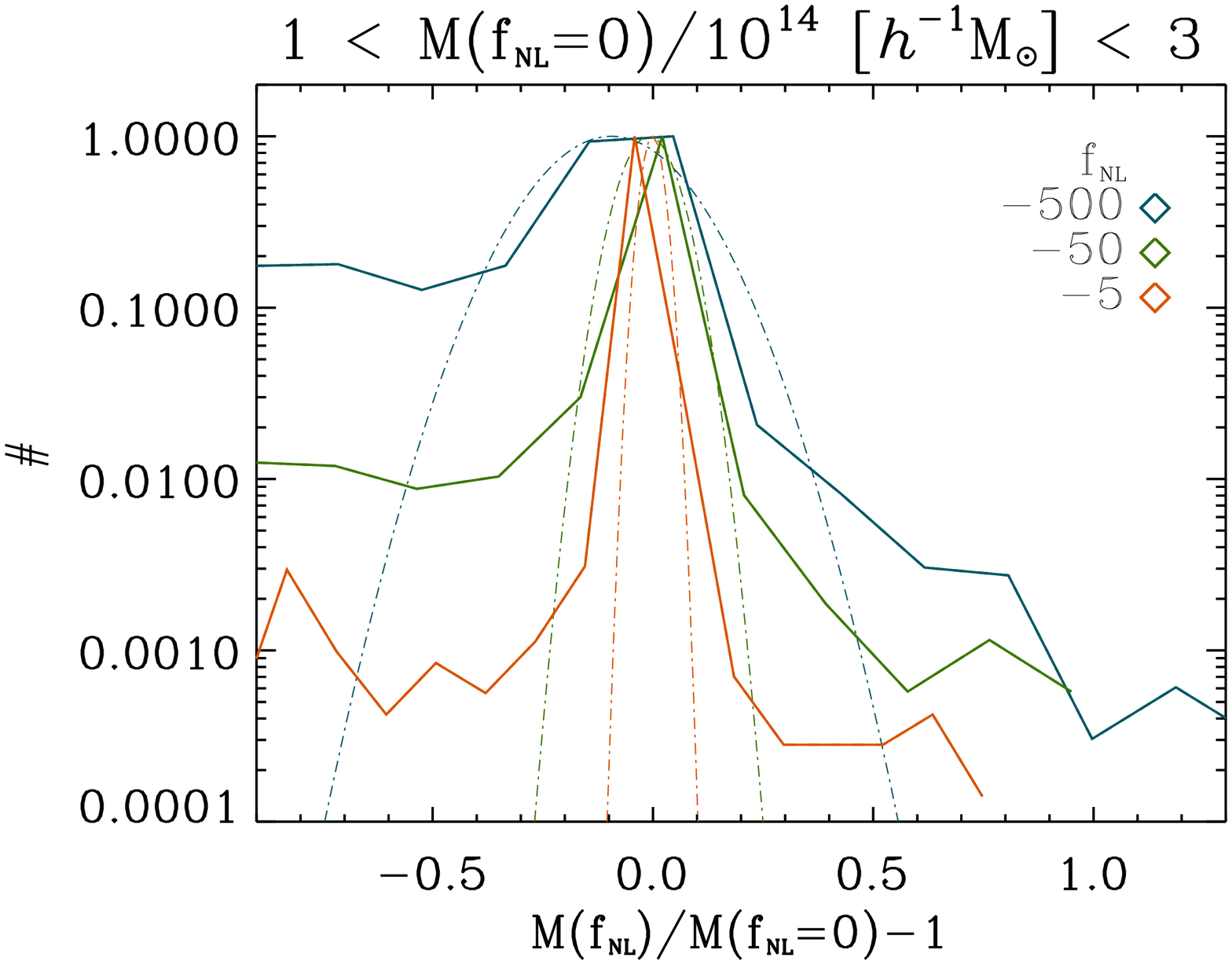,width=0.36\textwidth,angle=+90}
\caption{The probability distribution that a Gaussian
        halo of mass $M_0$ maps to a non-Gaussian halo of mass $M_f$, i.e.\
        $dP/dM_f(M_0)$. This plot can be understood as a (binned)
        slice through Fig.~\ref{fig:scatter_momf}.
        Here we show the measured $dP/dM_f(M_0)$ (solid lines)
        and Gaussian fit (dashed line) for various $\fnl$ in the mass bin
        $1<M_0/10^{14} M_{\sun}<3$. The left panel corresponds to $\fnl>0$ and
        the right panel corresponds to $\fnl<0$.  Note that both the width and
        mean value of the PDF vary with $\fnl$.  The probability distribution
        is clearly poorly fit by a Gaussian, however as discussed in the text,
        it provides an adequate fit given the precision with which we can
        determine the halo mass function from N-body
        simulations.  Whereas the high mass tail for $\fnl>0$ (left
        panel) indicates that many $\fnl=0$ halos will merge into more massive
        ones, the low mass tail for $\fnl<0$ (right panel) accounts for the
        disruption of $\fnl=0$ halos into lighter ones.
\label{fig:pdf_compare}}
\end{figure*}

\begin{figure*}[!t]
\epsfig{file=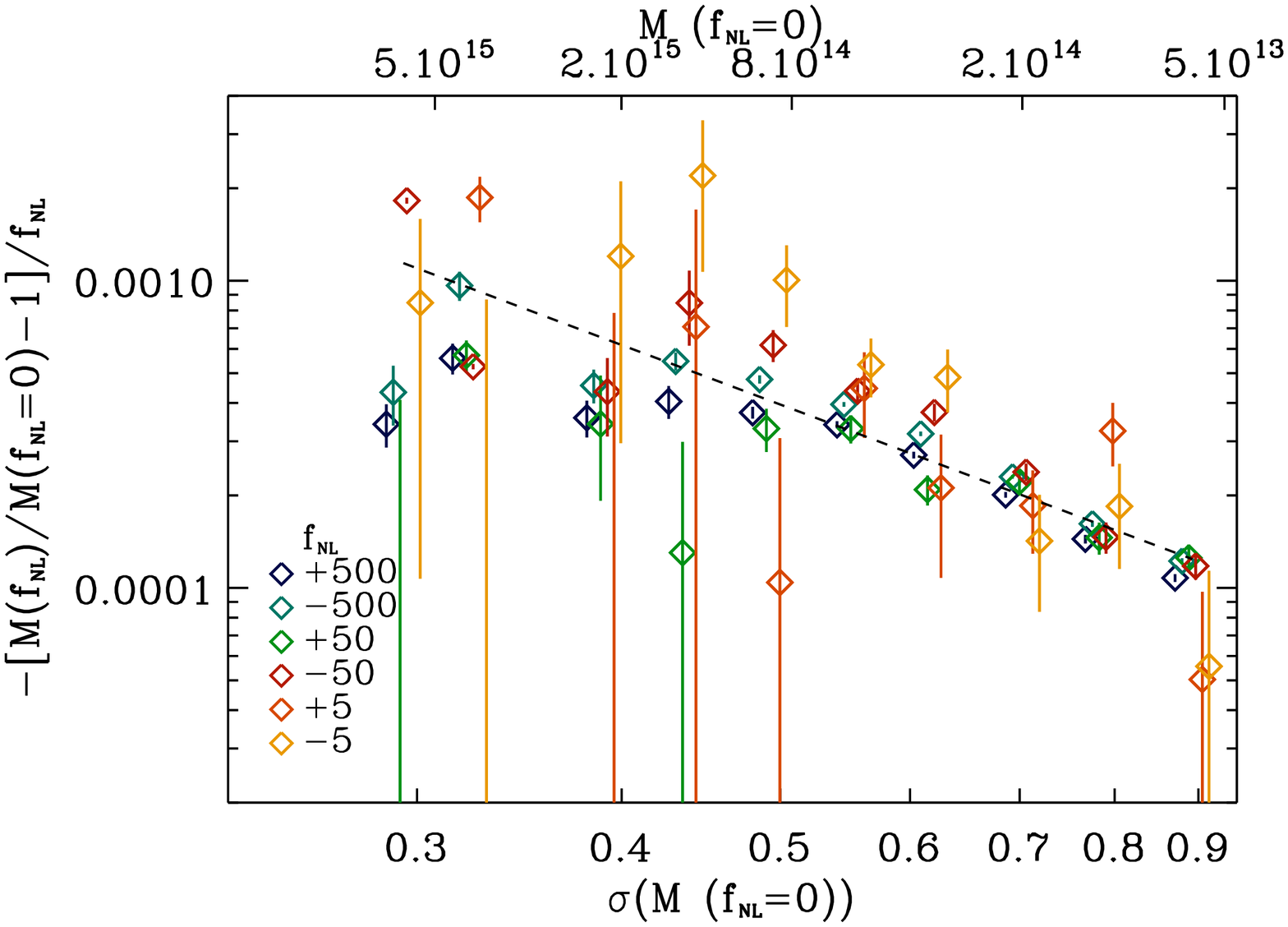,width=0.35\textwidth,angle=90}
\epsfig{file=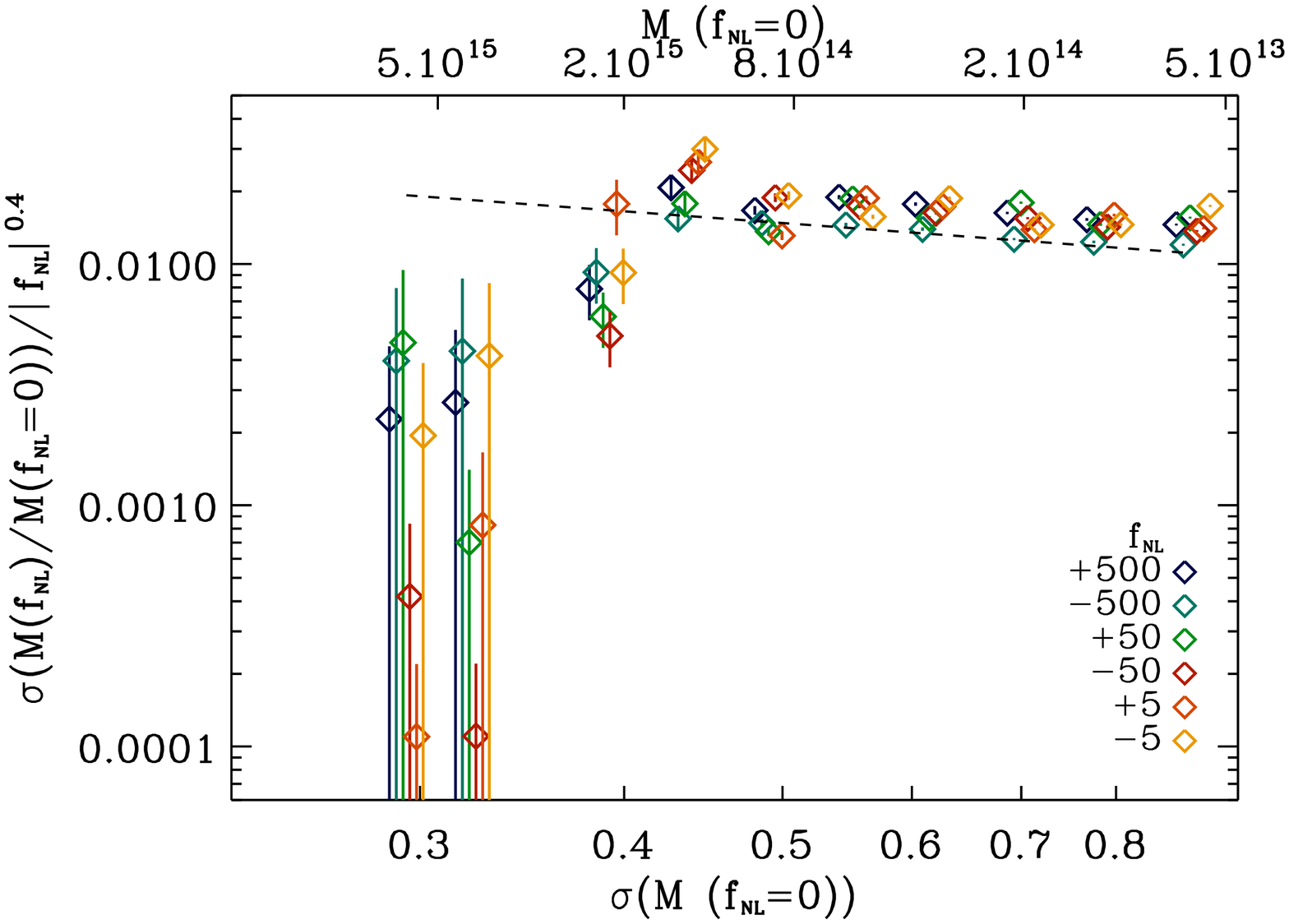,width=0.35\textwidth,angle=90}
\caption{Measured mean (left) and rms dispersion (right) of the mass
shift, $M_f/M_0 - 1$, as a function of $\sigma(M_0,z)$.  Note
that measurements at various redshift outputs ($z=0,0.5,1$) have been
combined in this plot, and that the $\fnl$ scaling has been
divided out.  The dashed lines shows our fits to these moments,
c.f. Eqs.\ (\ref{eq:mean_Mf_def}) and (\ref{eq:rms_Mf_def}).
\label{fig:scatter_mean_rms}}
\end{figure*}

Examples of the probability distribution are shown in Figure~\ref{fig:pdf_compare}, and
the mean and variance of the PDF are plotted in Figure~\ref{fig:scatter_mean_rms}.  The
behavior of the mean $\langle M_f\rangle$ and variance are quite regular,
and appear consistent with simple power laws:
\bea
\left\langle\frac{M_f}{M_0}\right\rangle -1
& = & 1.3\ 10^{-4}\,\fnl \sigma_8\,\sigma(M_0,z)^{-2} \label{eq:mean_Mf_def}
\\[0.1cm]
{\rm var}\left(\frac{M_f}{M_0}\right)
& = & 1.4\ 10^{-4}\, (\fnl\sigma_8)^{0.8} \sigma(M_0,z)^{-1},
\label{eq:rms_Mf_def}
\eea
where the rms overdensity dispersion $\sigma(M,z)$ is defined as usual by 
\begin{equation}
\sigma^2 = \int \frac{k^3}{2\pi^2} P(k)\,W^2(kR) \frac{dk}{k},
\label{sigma}
\end{equation}
where we use a top-hat window $W(x)=3 j_1(x)/x$ for 
$R=(3M/4\pi\bar\rho_m)^{1/3}$, and $P(k)$ and $\bar\rho_m$ are the matter
power spectrum and energy density  respectively.

Because we desire a simple fitting formula, we assume that we can approximate
the PDF as a normalized Gaussian whose mean and variance are given above, even
though the PDF shape is quite clearly nongaussian (c.f.\
Figure~\ref{fig:pdf_compare}).  As we show below, however, even this crude
approximation is sufficient to achieve the $\sim 10\%$ precision in the halo
mass function provided by standard fitting formulae for Gaussian simulations
\cite{lukic}.  Then Eq.~(\ref{eq:mf_conv}), together with $dP/dM_f(M_0)$ which is
assumed to be a Gaussian with the mean and variance given in
Eqs.~(\ref{eq:mean_Mf_def}) and (\ref{eq:rms_Mf_def}), fully specify our fitting
function.  Essentially, we have written the NG mass function as a convolution
of the Gaussian mass function with a Gaussian kernel.


\subsection{Review of previous fitting formulae}\label{sec:EPS_MVJ}

Before showing a comparison of the simulated mass function
to our proposed fitting formula, we first describe
alternative fitting formulae previously suggested in the literature: 
the Extended Press-Schechter (EPS) formalism \cite{Sefusatti}, and
the model of \citet[hereafter MVJ]{MVJ}.

\subsubsection{Extended Press-Schechter}

The EPS formalism generalizes the widely used Press-Schechter
\cite{press-schechter} model, which posits that the fraction of mass
in collapsed objects is equal to {\it twice} the fraction of the
volume occupied by density peaks exceeding some critical 
overdensity $\delta_c$. (The factor of two arises from the so-called
`cloud-in-cloud' problem \cite{bcek}.)  Therefore the collapsed
fraction becomes
\begin{eqnarray}
F(>M)&\equiv&2\int_{\delta_c}^{\infty} P(\delta;M) d\delta\label{eq:F(M)}\\
&=&2\,\int_{\delta_c/\sigma(M)}^{\infty} P_{\rm G}(\nu)d\nu
\end{eqnarray}
where the probability distribution $P(\delta; M)$ that the density
smoothed on mass scale $M$ equals $\delta$ is simply $P_{\rm G}(\nu)$,
the Gaussian distribution with zero mean and unit variance for
$\nu=\delta_c/\sigma$, with $\sigma(M)$ given by Eq.~(\ref{sigma}).
Note that $F(0)=1$ for hierarchical cosmologies where $\sigma(M)$
diverges as $M\rightarrow 0$; that is, {\em all} matter is assumed to
be in virialized objects of some mass.

The differential mass function may then be readily computed:
\begin{equation}
M\frac{dn}{dM} = {\bar\rho}_m \left |{dF\over dM}\right |.
\label{eq:dndM_F(M)}
\end{equation}
For a Gaussian PDF, the mass $M$ enters the right-hand side of this
expression only via the lower bound of the integral,
$\delta_c/\sigma(M)$, so we immediately obtain
\begin{equation}
\left ({dn\over d\ln M}\right )_{\rm PS} 
= 2\,{{\bar\rho}_m\over M}{\delta_c\over \sigma}\left
|{d\ln\sigma\over d\ln M}\right | P_{\rm G}(\delta/\sigma).
\label{eq:PS}
\end{equation}
This gives the well-known Press-Schechter mass function.  

\begin{figure*}[!t]
\epsfig{file=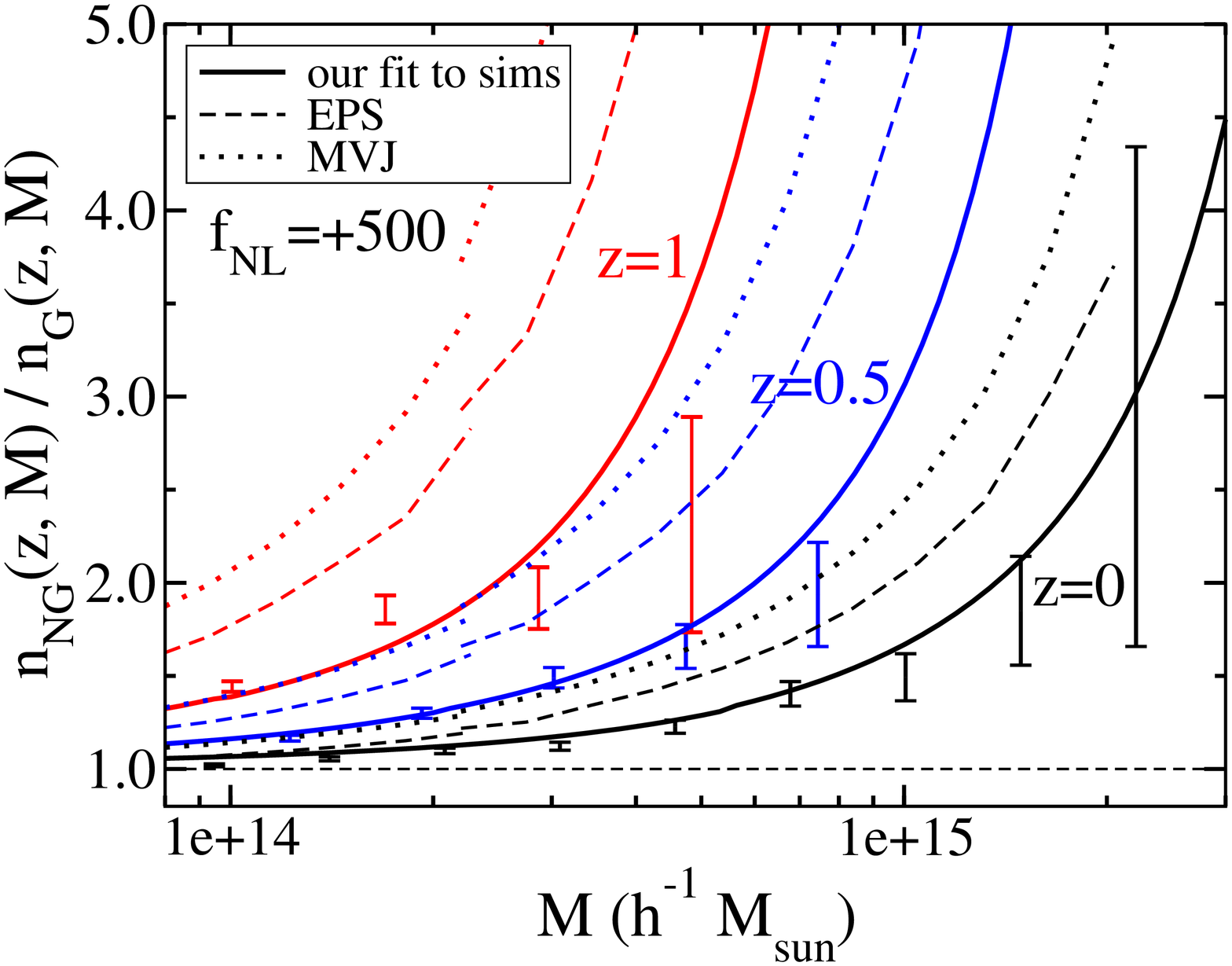,height=3.4in,width=2.8in,angle=-90}
\epsfig{file=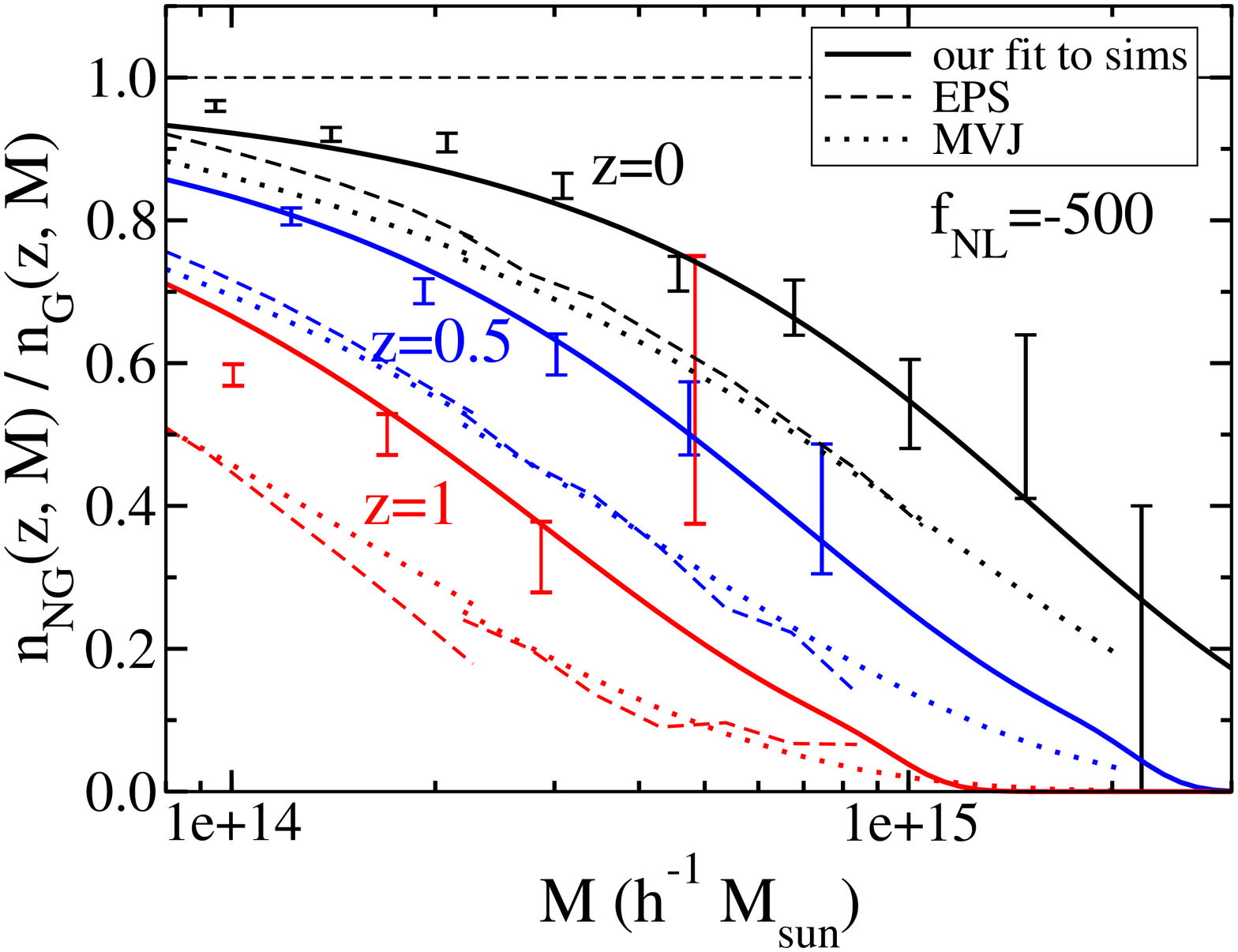,height=3.4in,width=2.8in,angle=-90}
\caption{Ratios of the NG to Gaussian mass functions as a function of mass and
at redshifts $z=0$ (black), $z=0.5$ (blue) and $z=1$ (red).  Points with error
bars denote results from our simulations. Solid lines represent our fitting
formula. Dashed and dotted lines denote the EPS
and MVJ fitting functions respectively. Note that the EPS and MVJ agree
mutually, but both significantly overestimate the effects of nongaussianity.
(The discontinuity of EPS and MVJ fitting functions at $M\sim 2\cdot 10^{14}\,
M_\Sun$ is due to transition from a smaller simulation box to the larger one.)
}
\label{fig:mf_compare_abs}
\end{figure*}

A class of fitting formulae based on this approach, and loosely called
`Extended Press-Schechter' (though apparently unrelated to the work of Refs.\
\cite{laceycole,bower}) attempt to generalize this argument using nongaussian
PDFs.  The most obvious way of making this generalization, i.e.\ inserting the
nongaussian PDF $P(\delta;M)$ into Eq.~(\ref{eq:dndM_F(M)}), faces several
immediate difficulties, however.  First, the Press-Schechter factor of 2 is no
longer valid; rather the cloud-in-cloud correction will depend upon the
specific form of nongaussianity.  Second, the shape of the PDF $P(\delta; M)$
now depends upon $M$ and so we cannot simply replace the derivative of the
integral in Eq.~(\ref{eq:dndM_F(M)}) by the integrand.  Lastly, and
prosaically, the Press-Schechter mass function does not, in fact, fit the halo
mass function in N-body simulations well, and so starting from PS is guaranteed
to fail in fitting the nongaussian mass function.

The approach adopted by many previous workers (e.g.\ \cite{Sefusatti})
has instead been to assume that, although Press-Schechter cannot be
used to derive the Gaussian mass function, it may be used to compute
the departure of the mass function from its Gaussian value, i.e.
\begin{equation}
\frac{n_{\rm NG}(M,z)}{n_{\rm G}(M,z)} = 
{{\ds d\over dM}\,F_{\rm NG}(>M) \over {\ds d\over dM}\,F_{\rm G}(>M)}
\label{eq:ratio}
\end{equation}
where $F$ is given by Eq.~(\ref{eq:F(M)}).  In this approach, the
nongaussian mass function is computed by multiplying the Gaussian mass
function (not Press-Schechter, but \citet{jenkins} or \citet{warren})
by the above ratio.

The EPS prediction for the halo mass function is therefore given by the
derivatives of the PDF tails given in Eq.~(\ref{eq:ratio}) above. To implement
this prescription, we compute the PDF tails directly from the initial
conditions of our simulations: at each redshift we are interested in, we
integrate the linearly evolved PDF to compute $F_{\rm NG}(M)$ at masses ranging
from about $ 10^{12}$ to $10^{16}M_\Sun$.  Then we compute the mean value of
$F_{\rm NG}(M)$ averaged over 10 independent N-body simulations. Finally, we
fit a cubic spline through the (computed mean of) $F_{\rm NG}(\ln M)$ and
differentiate with respect to mass.  Evaluation of this formula
becomes extremely difficult at high masses, simply because the
statistics of peaks at these high masses becomes too noisy.  

\subsubsection{MVJ}

The MVJ \cite{MVJ} mass function is a further approximation to the EPS
model described above.  Instead of numerically computing the PDF and
its tails for each $\fnl$, it is assumed that the ratio in
Eq.~(\ref{eq:ratio}) may be determined from the skewness of the PDF.
The expression for the mass function becomes \cite{Grossi}

\begin{eqnarray}
\left ({dn\over d\ln M}\right )_{\rm MVJ} &=& 
2{\bar\rho_M\over M} P_G\left ({\delta_*\over \sigma_M}\right ) \times\\[0.1cm]
&&\left [{1\over 6}{\delta_*^3\over \delta_c}\left |{dS_{3, M}\over d\ln M}\right | +
  \delta_* \left |{d\sigma_M\over d\ln M}\right | \right ]\nonumber
\label{eq:MVJ}
\end{eqnarray}
where
\begin{eqnarray}
\delta_* &\equiv& {\delta_c\over \sqrt{1-S_{3, M}\delta_c/3}}\\[0.1cm]
S_{3, M} &\equiv & {\langle \delta^3\rangle_M\over \langle\delta^2\rangle_M^2}
\end{eqnarray}
where $\langle \delta^n\rangle_M$ is $n$th moment of the density field
evaluated on the characteristic mass scale $M$, and $S_{3, M}$ is the
skewness on that mass scale.  

The advantage of the MVJ formula is that it does not require
specification of the PDF of the density field --- however, it does
require knowledge of the skewness $S_{3, M}$.  In this work, we compute the
moments of the density field directly from our simulations at the
starting epoch $a=0.02$, then scale them with the linear growth
function to the desired epoch. We can then evaluate the MVJ expression
in Eq.~(\ref{eq:MVJ}).  Unlike the EPS approach described above, the
formula does not become intractable at high masses.  On the other
hand, MVJ becomes onerously expensive to calculate at 
low levels of nongaussianity (e.g.\ $|\fnl|\lesssim100$), simply
because the scatter in the measured skewness from run to run becomes
comparable to that generated by primordial nongaussianity.  To see
this, note that the scatter in $\langle\delta^3\rangle_M$ is roughly
$\sigma_3\simeq\sigma_M^3 \sqrt{15/N}$ for $N=M_{\rm box}/M$ samples.
Approximating $\langle\delta^3\rangle_M\sim 6\fnl\sigma_M^3\sigma_\phi$, 
then for a $512^3$ grid and mass scale of 200 cells, and taking
$\sigma_\phi=4\times10^{-5}$, requiring 
$\langle\delta^3\rangle_M/\sigma_3>5$ translates into $|\fnl|\gtrsim 100$.

\subsection{Results and comparison to previous work}
\label{sec:comparison}

Figure \ref{fig:mf_compare_abs} shows the ratios $n_{\rm NG}/n_{\rm G}$ for
$\fnl=500$ (left panel) and $-500$ (right panel). Simulation values are denoted
with error bars, colored black ($z=0$), blue ($z=0.5$) and red ($z=1$). To
compute the error bars in the ratios, and taking account of the fact that
$n_{\rm NG}$ and $n_{\rm G}$ measurements are correlated, we have adopted the
larger error of the two alone (rather than adding them in quadrature), which is
the error in $n_{\rm G}$ ($n_{\rm NG}$) for $\fnl>0$ ($\fnl<0$).  The solid
lines denote our fits explained in Sec.~\ref{sec:method}. Dashed lines refer to
the EPS results, while the dotted lines represent the MVJ fitting function.
The results clearly indicate that, while the EPS and the MVJ functions mutually
agree\footnote{The agreement between the EPS and MVJ is even better when an
alternative expression is used in for $\fnl>0$, as pointed out
\citet{Grossi}; see their Eq.~(4). We have not used this correction in
our 
Fig.~\ref{fig:mf_compare_abs}.}, they both overestimate the effects of
nongaussianity as found by our simulations, at a level typically $\lesssim
100\%$ although dependent upon mass and redshift.

This result appears to disagree with the work of \citet{Kang}, who find a large
discrepancy between EPS/MVJ and their simulations' mass function, in the sense
that their simulations show a much {\em larger} effect of nongaussianity than
predicted by the EPS type formalism.  However, as noted by these authors, their
simulations used a rather small number of particles ($\sim 128^3$) in a volume
nearly 20$\times$ smaller than ours, so it is unclear how well they probe the
statistics of the rare objects of interest to us.  In contrast, \citet{Grossi}
have found very good agreement between the MVJ formula and their simulations'
results.  While our fitting function is in mild disagreement with the MVJ
fitting formula, it is unclear whether our simulations are in disagreement with
the simulations of \citet{Grossi}.  Their simulations used a somewhat different
cosmological model (higher $\sigma_8$) than ours, they have plotted cumulative
rather than differential mass functions, and of course the error bars in both
their plots and ours are considerable.

In summary, we conclude that our simple fitting function appears consistent
with the measured mass function from our simulations to within $\sim10\%$ over
the entire range of masses and redshifts that we consider.  Since this is the
level of precision that various N-body codes agree with each other in the mass
function \cite{lukic}, we have not attempted to achieve better agreement.
EPS-like fitting formulae, such as the model of MVJ \cite{MVJ}, appear to
overestimate the effects of nongaussianity.  The level of discrepancy increases
with increasing mass and redshift.

\section{Halo clustering} \label{sec:pk}

\begin{figure}[!t]
\epsfig{file= 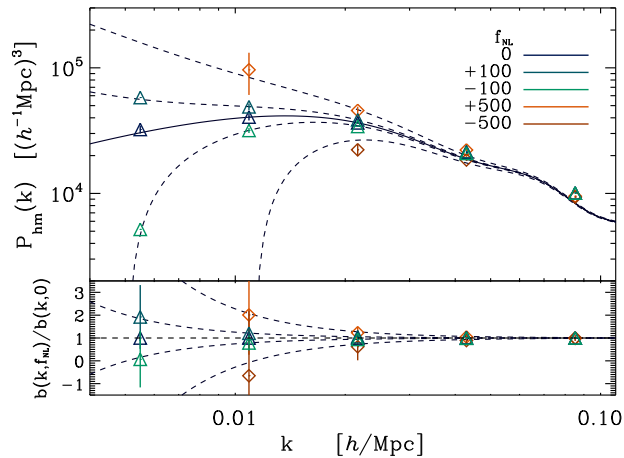,angle=+90,width=0.5\textwidth}
\caption{Cross-power spectra for various $\fnl$.  The upper panel
  displays $P_{h\delta}(k)$, measured in our simulations at $z=1$ for
  halos of mass $1.6\times 10^{13} M_{\Sun} < M < 3.2\times 10^{13}
  M_{\Sun}$. The solid line corresponds to the theoretical prediction
  for $P_{\delta\delta}$ with a fitted bias $b_0$=3.25. We see a
  strongly scale-dependant correction to the bias for $\fnl \neq 0$,
  increasing towards small $k$ (large scales). The bottom panel
  displays the ratio $b(k,\fnl)/b(k,\fnl=0)$.  The errors are computed
  from the scatter amongst our simulations and within the
  bins. Triangles correspond to our large ($1024^3$ particle)
  simulations whereas diamonds correspond to our smaller ($512^3$
  particle) simulations. The dotted lines correspond to our expression for
  the bias dependence on $\fnl$ defined in Eq.~(\ref{bias_eul}).}
\label{fig:plot_pk_bk}
\end{figure}

Beyond one-point statistics like the halo mass function, N-body
simulations also allow us to compute higher order statistics like the
correlation function or its Fourier transform, the power spectrum.  
As shown in Sec.~\ref{NG_analytic}, we expect nongaussianity to
produce pronounced effects on the halo power spectrum, specifically in
the form of scale-dependent halo bias on large scales.  This may seem
somewhat surprising, due to very general arguments previously given in
the literature that galaxy bias is expected to be independent of scale
in the linear regime \cite{coles93,fry93,scherrer98}.  We can summarize
the argument as follows.  Suppose that the halo overdensity is some
deterministic function of the local matter overdensity, $\delta_h =
F(\delta)$.  On large scales, where $|\delta|\ll 1$, we can Taylor
expand this function, $\delta_h = a + b\,\delta + \ldots$.  Keeping
only the lowest order terms and requiring that
$\langle\delta_h\rangle=0$ then gives $\delta_h = b\,\delta$, which is
linear deterministic bias.  The key assumption in this argument was
locality; i.e. that the halo abundance is determined entirely by the
local matter density.  N-body simulations with Gaussian initial
conditions have confirmed that halo bias tends to a constant on large
scales well in the linear regime.

Once we allow for primordial nongaussianity, however, the above
argument need not hold.  For example, in this paper we have considered
NG of the form $\fnl\Phi^2$, and note that the gravitational potential
is a nonlocal quantity.  Hence the locality-based argument above does
not apply for this form of nongaussianity, and our derived
scale-dependence of the bias is not surprising.  The specific form we
have derived is particular to the quadratic, local form of NG that we
have assumed, however we expect any NG that couples density modes with
potential modes will in general lead to scale-dependent bias.  On the
other hand, nongaussianity of the form $\fnl\delta^2$ does not
lead to scale-dependent bias.  

In order to test our prediction for the scale dependence of bias, 
we have computed halo bias in our N-body simulations by taking the
ratio of the matter power spectrum $P_{\delta\delta}$ and the
halo-matter cross spectrum $P_{h\delta}=\langle\delta_h^*\delta\rangle$.
We have used the cross spectrum 
rather than the halo auto spectrum because the former should be less
sensitive to shot noise from the small number of halos compared to DM
particles.  We have checked, however, that using the halo auto-spectra
to compute bias gives consistent results as the cross-spectra; i.e.\
we find no evidence for stochasticity.  Examples of the various power
spectra and resulting bias factors are plotted in figure
Fig.~\ref{fig:plot_pk_bk}.  

\begin{figure}[!t]
\epsfig{file= 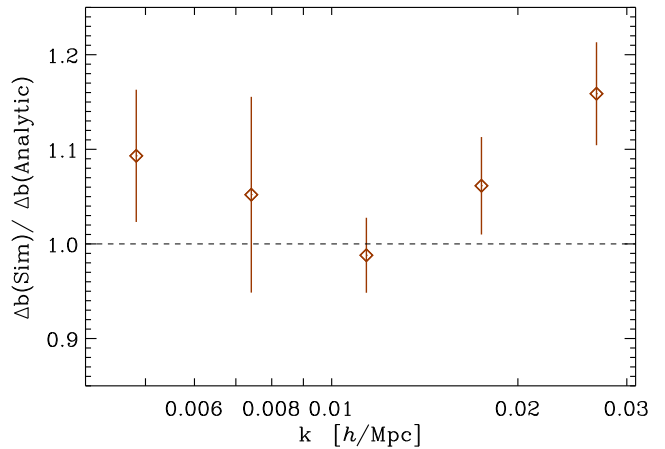,angle=+90,width=0.5\textwidth}
\caption{Ratio of the bias shift $\Delta b$ measured from our
  simulations to that predicted by Eqn.~(\ref{bias_eul}), using
  $\delta_c=1.686$.  Biases were computed from cross-spectra
  measured on 28 simulations with 5 various $\fnl$ (-500, -100, 100,
  500), 3 various redshifts ($z=0,0.5,1$) and 5 halo mass bins. Note
  that at higher $k$, nonlinear evolution also  
  generates scale dependence in the bias \cite{smith07a}.}
\label{fig:plot_bk_scatter}
\end{figure}

As can be seen, we numerically confirm the form of the predicted scale
dependence.  Because we focus on the statistics of rare objects, the
errors on bias from individual simulations plotted in
Fig.~\ref{fig:plot_bk_scatter} is large.  We therefore attempt to
improve the statistics on the comparison by combining the bias
measurements from multiple simulations.  Figure
\ref{fig:plot_bk_scatter} plots the average ratio between the bias
measured in our simulations and our analytic prediction
Eqn.~(\ref{bias_eul}), using $\delta_c=1.686$ as predicted from the
spherical collapse model \cite{GunnGott72}.  In computing the average
plotted in this figure, we used a uniform weighting across the
different simulations, redshifts, and mass bins.  Alternative
weightings can shift the results by $\sim 10\%$, so we conservatively
estimate the systematic error in our comparison to be 20\%.  
The agreement between our numerical
simulation results and our predicted bias scale-dependence,
Eqn.~(\ref{bias_eul}), is excellent and perhaps surprising.  Naively,
we might expect a somewhat larger collapse threshold $\delta_c$ to
apply, considering the ellipsoidal rather than spherical nature of the
collapse of halos in this mass range \cite{bm96}.

\section{Cosmological consequences} \label{sec:cosmo}

Having derived fitting formulae for the abundance and clustering of
halos in NG models, we now investigate how well upcoming surveys may
constrain $\fnl$, and whether NG could possibly affect the constraints
derived on other cosmological parameters.  We focus on galaxy cluster
surveys and redshift surveys.  Cluster surveys aim to constrain
cosmological parameters, in particular dark energy parameters, by
exploiting the exponential sensitivity of the galaxy cluster abundance
on cosmology.  Similarly, a major goal for upcoming redshift surveys
is to constrain dark energy by localizing baryonic acoustic
oscillation (BAO) features in the galaxy power spectrum at multiple
redshifts.  Examples of upcoming surveys include the 
Atacama Cosmology Telescope\footnote{http://wwwphy.princeton.edu/act/}, 
South Pole Telescope\footnote{http://spt.uchicago.edu},
Dark Energy Survey\footnote{http://www.darkenergysurvey.org},
WiggleZ\footnote{http://astronomy.swin.edu.au/wigglez/WiggleZ/Welcome.html},
Planck\footnote{http://www.rssd.esa.int/Planck},
SuperNova/Acceleration Probe\footnote{http://snap.lbl.gov},
and the Large Synoptic Survey Telescope\footnote{http://www.lsst.org}.

Because primordial nongaussianity affects both the abundance and power
spectra of massive halos, both of these types of surveys will be
well-suited for constraining NG.  On the other hand, potential NG
could in principle degrade the expected constraints on dark energy
parameters, due to possible degeneracies.  We use the Fisher matrix
formalism to extract errors on seven cosmological parameters as well
as $\fnl$.  Our estimates are only illustrative; accurate forecasts for
specific surveys will require a more sophisticated analysis.

\subsection{Constraints from $P(k)$:\\ Galaxy surveys, BAO and ISW}

We can (crudely) estimate constraints on parameters $\{p_i\}$ derived
from measurements of the power spectrum by assuming that bandpowers
are measured with errors $\delta P = (P + n^{-1})/\sqrt{m}$, where $P$
is the power in a band of width $dk$ centered at wavenumber $k$, $n$
is the number density of galaxies, and $m$ is the number of
independent Fourier modes sampled by the survey, roughly given by
$m=(2\pi^2)^{-1} V\, k^2 dk$ \cite{blake06}.  
Then the Fisher matrix can be written as
\begin{equation}
F_{ij}=\int_{k_{\rm min}}^{k_{\rm max}}\frac{\partial P}{\partial p_i}
\frac{\partial P}{\partial p_j}\left(P+\frac{1}{n}\right)^{-2}
\frac{V\,k^3}{2\pi^2} d\ln k.
\label{fishy_pk}
\end{equation}
For simplicity, in Eq.~(\ref{fishy_pk}) we use the linear theory power
spectrum and 
number density corresponding to $z=0.5$, and assume an all-sky,
volume-limited survey extending to $z=0.7$.  We integrate over
wavenumbers between $k_{\rm min}=10^{-3} h/$Mpc and $k_{\rm max}=0.1
h/$Mpc.  The results are insensitive to $k_{\rm max}$ but depend
strongly on $k_{\rm min}$.  We believe the $k_{\rm min}$ used here is
optimistic but reasonable.  At high $k$ (small scales), late-time
nonlinear evolution can also generate scale-dependent bias
\cite{smith07a}, however the redshift and scale dependence of this
effect is quite distinctive from NG and we ignore it here.

We assume that the target galaxies have properties similar to luminous
red galaxies (LRGs) \cite{LRG07}, with comoving number density
$n=4\times 10^{-4} (h^{-1}{\rm Mpc})^{-3}$ and bias $b_0=2$.
Equation~(\ref{fishy_pk}) then gives estimated errors on $\fnl$ of
$\sigma(\fnl)\approx 7$, which compares well with forecasted
constraints on nongaussianity for Planck. 

Unsurprisingly, we find little degeneracy between $\fnl$ and other
cosmological parameters, given its distinctive effect on the shape of the power
spectrum.  Accordingly, there is little reason to believe that BAO
determinations of dark energy parameters will be biased by nongaussianity,
especially since the scale dependence is small over the wavenumbers of interest
for the BAO wiggles. To quantify this effect we determine the acoustic
peak position by looking at extrema of the ratio of the power spectra
with baryons and the power spectrum with zero baryons
\cite{Hu_transfer}. When multiplying the matter power spectrum with
baryons by our scale dependant bias, we find that $\fnl=100$ would shift
the first BAO peak at $k\simeq 0.07 h/$Mpc by 0.4\% at $z=1$, and has
a considerably smaller effect at the higher BAO peaks.  The magnitude
of this effect is comparable to the effect of non-linear corrections
to the power spectrum \cite{Crocce:2007dt,smith07b}, although the NG effect is
primarily important on large scales while nonlinearities are most
important on small scales.  In principle, NG and nonlinearities could
conspire to lead to a $\sim1-2\%$ bias in the dark energy equation of
state parameter $w$ inferred from BAO observations
\cite{Crocce:2007dt}, and so a careful joint analysis allowing both
for NG and nonlinear corrections will be required, which should not be difficult. 

Another probe of $P_{h\delta}$ on large scales is the cross-correlation
between cosmic microwave background (CMB) temperature anisotropies and 
large scale structure, due to the integrated Sachs-Wolfe (ISW) effect
\cite{Sachs:1967er,Crittenden:1995ak,Bean:2003fb}. First detections of the ISW
effect from cross-correlations of WMAP with various large scale surveys have
been obtained with reported detections at the 2-4$\sigma$ level
\cite{boughn04,nolta04,fosalba03,scranton03,fosalba04,Padmanabhan:2004fy,afshordi04a,cabre06}. 
A combined analysis yields a $\simeq 5 \sigma$ detection
\cite{scranton07}, whereas a cosmic variance limited measurement would
allow a $\simeq 7.5\sigma$ detection for the currently favored
$\Lambda$CDM cosmology, and a somewhat more
significant detection if the dark energy equation of state parameter
is smaller \cite{Afshordi:2004kz,Hu:2004yd}. The cross-correlation 
between large-scale structure and CMB is directly proportional to a weighted
projection of the scale-dependant bias. Since the $z$ and $k$ dependence of our
bias is very specific, we do not expect it to be severely degenerate with other
parameters affecting the amplitude of the ISW effect (mostly $w$ and $\Omega_m$
for a flat universe). We can thus translate the ISW detection level into
constraints on $\fnl$. For the sake of simplicity, we assume that the ISW
signal comes from $z\simeq 1$ and is dominated by the angular multipole
$\ell\simeq 20$, corresponding to a wavenumber $k\simeq 6.66\times 10^{-3} h/$Mpc at $z=1.$
\cite{Afshordi:2004kz}.  According to Eq.~(\ref{bias_eul}) the current
$3\sigma$ ($5\sigma$) detections of ISW translate into upper limits on $|\fnl|$
of 123 (61) (1$\sigma$) assuming a bias $b_0=2$, as appropriate for LRGs
\cite{Padmanabhan:2004fy}. A prospective 7.5 $\sigma$ detection would translate
into $|\fnl|\lesssim 38$ (1$\sigma$).  These estimates are clearly very crude, but
are likely correct at the order of magnitude level.  In comparison,
the current limit from CMB bispectrum measurements from WMAP give
-54 $< \fnl <$ 114 (95\% CL) \cite{wmap3} whereas Planck is expected to
constrain $|\fnl|<10$ (1 $\sigma$) \cite{Smith:2006ud}.

In summary, the large-scale galaxy power spectrum appears capable of
constraining local NG quite stringently for surveys reaching $\sim$Gpc
scales: $|\fnl|\lesssim 10$.  ISW or BAO observations could 
provide somewhat weaker bounds on $\fnl$, though of course any
constraints they can provide would be independent of the CMB
bispectrum and therefore worthwhile.  Our estimates of forecasted
bounds on $\fnl$ were rather crude, but given the encouraging results,
a more sophisticated treatment for specific survey parameters appears
warranted.

\subsection{Constraints from cluster counts}

We next consider how well upcoming cluster surveys can constrain
$\fnl$ by measurements of the cluster mass function $dn/dM$.  Other
forms of nongaussianity may also be constrained by these surveys
\cite{sadeh07}, but we focus on the $\fnl$ form. For our
fiducial survey parameters, we consider a fixed, redshift-independent lower 
mass limit of $M_{\rm lim}=2\times 10^{14}M_{\sun}$ and assume redshift bins
of width $\Delta z=0.1$ uniformly distributed between $z=0.1$ and $z=\zmax\leq 2.0$.  
We simultaneously vary seven cosmological parameters besides $\fnl$:
$A$, the normalization of the primordial power spectrum at 
$k_{\rm fid}=0.002\hmpcinv$; physical matter and baryon densities
$\Omega_m h^2$ and $\Omega_b h^2$, spectral index $n_s$, the sum 
of the neutrino masses $m_{\nu}$, the matter energy density today relative to
critical $\Omega_m$, and the equation of state parameter of dark energy $w$.
We assume no mass information
(which would improve our parameter constraints) but also no systematic errors
(which would degrade the constraints). We further assume 5000
square degrees on the sky, roughly consistent with expectations for the Dark
Energy Survey or the South Pole Telescope. The fiducial survey has about 7000
clusters (for $\sigma_8=0.76$ cosmology) and about 23,000 for $\sigma_8=0.9$).
We use WMAP3 \cite{wmap3} cosmological parameters in determining error
forecasts.  The mass power spectrum $\Delta^2(k, a)\equiv k^3 P(k,
a)/(2\pi^2)$ is written as

\begin{equation}
\Delta^2(k, a) = \frac{4A}{25\Omega_M^2} 
\left ({k\over k_{\rm fid}}\right )^{n_s-1}
\left ({k\over H_0  }\right )^4
g^2(a)\,T^2(k)
\label{eq:P(k)}
\end{equation}

\noindent where $T(k)$ is the transfer function adopted from \citet{Hu_transfer}, 
and the growth function $g(a)$ is computed exactly by
integrating the well known second order differential equation for growth (e.g.\
Eq.~(1) in \cite{Coo_Hut_Bau}). 

Following the results of section \ref{sec:fit_formula}, we assume that
the mass function may be written as
\begin{equation}
{dn\over dM}(z, M) = 
\left ({dn\over dM}\right )_{\hspace{-0.1cm}\rm Jenk}\hspace{-0.2cm} (z, M) \times
\left [ n_{\rm NG}(z, M)\over n_{\rm G}(z, M)\right ]
\end{equation}
where the nongaussian correction is computed using either our fitting
formula, or EPS for comparison.  For a
given mass function, the total number of objects in a redshift interval of
width $\Delta z$ and centered at $z$ is
\begin{equation}
N(z, \Delta z)=\Omega_{\rm survey}
\int_{z-\Delta z/2}^{z+\Delta z/2} n(z, M_{\rm min})\,{dV(z)\over d\Omega\,dz}\, dz
\end{equation}
\noindent where $\Omega_{\rm survey}$ is the total solid angle covered
by the survey, $n(z, M_{\rm min})$ is the comoving density of clusters
more massive than $M_{\rm min}$, and $dV/d\Omega dz$ is the comoving
volume element.  We assume $\Omega_{\rm survey}=5000$
square degrees, roughly consistent with expectations for the Dark
Energy Survey or the South Pole Telescope.  The fiducial survey has about 7000
clusters (for $\sigma_8=0.76$ cosmology) and about 23,000 (for $\sigma_8=0.9$).

Assuming Poisson statistics, the Fisher information matrix reads 
\cite{Holder_Haiman_Mohr, Huterer_Broderick, Lima_Hu_04}
\begin{equation}
F^{\rm clus}_{ij} = \sum_k {1\over N_k(z_k, \Delta z)} 
{\partial N_k\over \partial p_i}\,
{\partial N_k\over \partial p_j}
\end{equation}
where $p_i$ are the 8 cosmological parameters including $\fnl$, $N_k$ is the
number of clusters $k$th redshift bin, and the sum runs over the redshift bins
extending to maximal redshift $\zmax$ \footnote{Some attention needs to be
  paid when taking the derivative with respect to $\fnl$, as it is especially
  with this parameter that the assumption that the likelihood function
  Gaussian may be violated, leading to results that are weaker or stronger
  than a full likelihood calculation would reveal. We explore different values
  of $d\fnl$ and find convergence at $|d\fnl|\lesssim 30$. We also find,
  however, that the sensitivity to $\fnl$ (and also to the normalization $A$)
  is slightly higher when $d\fnl>0$ ($d\ln A>0$) than when $d\fnl<0$ ($d\ln
  A<0$); this is expected as excess of rare objects like galaxy clusters
  provides more cosmological leverage than their absence. This implies that
  true constraints on $\fnl$ will be slightly asymmetric around our fiducial
  value of zero; therefore, we make sure to take two-sided derivatives with
  $d\fnl=\pm 30$. For the EPS function, we take the two-sided derivative with
  $d\fnl=\pm 50$ (recall we ran sinulations with $|\fnl|=5$, $50$ and $500$),
  and check that the results are similar, if noisier, if $d\fnl=\pm 5$ is
  used.}. We assume no mass information (which would improve our parameter
constraints) but also no systematic errors (which would degrade the
constraints).

Lastly,
we add a Planck prior on the parameter set, neglecting forecasted
constraints on $\fnl$ expected from future CMB bispectrum
measurements (since we are interested in the sensitivity to NG of cluster
counts alone). The full Fisher matrix is given by 
\begin{equation}
F = F^{\rm clus} + F^{\rm CMB}.
\end{equation}

\begin{figure}[!t]
\epsfig{file=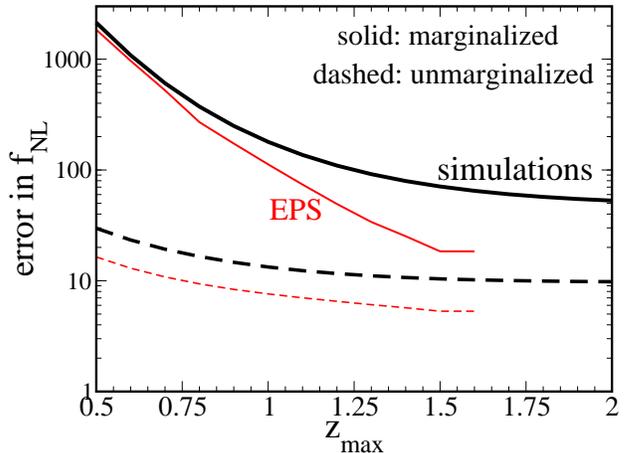,angle=-90,width=0.5\textwidth}
\caption{Forecasted errors on $\fnl$ from measurement of the cluster
mass function, as a function of the maximum extent of the
cluster survey, $\zmax$. Black solid and dashed line show the marginalized and
unmarginalized error using our fitting formulae for $n_{\rm NG}/n_{\rm G}$,
while the red lines show the errors using the EPS formalism. Even though we use
the average of the nongaussian PDF tail over 10 simulations, the EPS result
becomes noisy at $z\gtrsim 1.5$ due to poor sampling of the
tails. Furthermore, it is clear that the EPS errors on $\fnl$ underestimate
those based on our simulations by up to a factor of three.}
\label{fig:err_fnl_vs_zmax}
\end{figure}

Fig.~\ref{fig:err_fnl_vs_zmax} shows the result of our Fisher matrix
estimate, i.e.\ the forecasted errors on $\fnl$ as a function of the
maximum extent of the cluster survey, $\zmax$. Black solid and dashed
lines show the marginalized and unmarginalized error using our fitting
formulae for $n_{\rm NG}/n_{\rm G}$, while the red lines show the
errors using the EPS formalism. The former errors are clearly well
behaved, and asymptote at high $\zmax$ as expected since cluster
abundance rapidly vanishes. On the other hand EPS-produced $\fnl$
errors disagree with the simulations by up to a factor of
three. Moreover, even though we use the average of the nongaussian PDF
tail over 10 simulations, the EPS result becomes noisy at $z\gtrsim
1.5$ due to poor sampling of the tails.  The magnitude of the
discrepancy between EPS estimates and our simulations appears similar
even for the higher $\sigma_8=0.9$ model.  As with our estimates from
power spectrum constraints, we do not find significant degeneracies
between $\fnl$ and other cosmological parameters (correlation
coefficient $\lesssim 0.5$).

Since we find a weaker effect on cluster abundance than previous
formulae like EPS or MVJ, this implies that constraints
found by \citet{Sefusatti}, who performed a similar
Fisher matrix estimate but with somewhat different assumptions, will
be weaker once the NG sensitivity is calibrated off simulations.
Direct quantitative comparison to two other relevant papers, 
\citet{Kang} and \citet{Grossi}, is however
difficult since these authors do not compute cosmological
parameter error estimates.

\section{Discussion}

We have quantified the effects of primordial nongaussianity on the abundance
and power spectra of massive halos.  Our two principal results are as
follows. 

First, we have provided a new fitting formula for the halo mass
function. The formula is based on matching halos in Gaussian and non-Gaussian
simulations: for $\fnl>0$ the corresponding halos are more massive than in the
Gaussian case, and vice versa. The formula is consistent with the measured
mass function from our simulations to within $\sim10\%$ over the entire range
of masses and redshifts that we consider. Being essentially a convolution of
the Gaussian mass function and a Gaussian kernel
(Eqs.~(\ref{eq:mf_conv})-(\ref{eq:rms_Mf_def})), the formula is also easy to use
and does not require estimating the extreme tails of the nongaussian PDF of
the density field. Our results also indicate that previous work based on
Extended Press-Schechter type formulae overestimated the effects of
nongaussianity on the abundance of halos by a factor of $\sim 2$ over the
relevant mass scales.

Secondly, we showed both analytically and numerically that
nongaussianity (in the $\fnl$ model) leads to strong scale dependence
of the bias of dark matter halos.  We find remarkably good agreement
between our analytic expression and our numerical results.  
Measurement of the power spectrum
of biased objects therefore provides a new avenue to detect and measure
nongaussianity.  While cluster counts can constrain NG at a level
comparable to existing CMB constraints, $|\fnl|\lesssim 100$, we found that
future large-scale redshift surveys can potentially do much better,
roughly $|\fnl|\lesssim 10$.  We do not find significant degeneracies
between $\fnl$ and dark energy parameters in our Fisher matrix
calculations, either for mass function measurements or power spectrum
measurements.  More precise estimates will require considerably more
sophisticated treatments than we have attempted in our illustrative
examples above. 

We close this paper by considering, in light of our findings, the optimal
methods for constraining NG of the $\fnl$ form.  Measurements of the power
spectrum would appear the most promising; observations of high redshift, highly
clustered objects on large scales would allow the strongest constraints on the
scale-dependent bias signature of $\fnl$.  Fortunately, upcoming BAO surveys
will likely provide the necessary observations of, e.g. luminous red galaxies
(LRGs).  Photometric surveys may also be useful in this regard.  Since the
effects of NG are most pronounced on large scales, rather than small scales,
precise spectroscopic redshifts may not be necessary.  Photometric redshifts
with errors of order $\Delta z\approx 0.03$ have already been achieved for LRGs
and for optically selected groups and clusters with prominent red sequences
\cite{Padmanabhan:2004ic,Ilbert:2006dp,Yee:2007if}.  At $z=0.5$, this
corresponds to roughly 100 $h^{-1}$Mpc comoving, fairly small compared to the
$\sim$Gpc scales where NG becomes most important.  Since photometric surveys
can cover wider areas more deeply than spectroscopic surveys, they may turn out
to provide tighter bounds.  

Besides their abundance and clustering, the internal properties of
massive halos may also be sensitive to nongaussianity. 
For instance, the concentrations and substructure content of massive
halos have been found to depend upon primordial NG \cite{avila03}.
Our simulations lacked sufficient force resolution to explore this in
detail, but we note in passing that multiple groups find a tension
between observations of massive lensing clusters and theoretical
predictions for Gaussian perturbations
\cite{rcs,arcs04,hennawi07,broadhurst08}. 

Another intriguing possibility for probing primordial NG
is to use statistics of the largest
voids in the universe.  Just as the abundance and clustering of high
density peaks are affected by nongaussianity, so are the same
properties for deep voids (albeit with an opposite sign, c.f.\
Fig.~\ref{fig:slice_sims}).  In a sense, because voids are not as
nonlinear as overdense regions, their properties are more easily
related to the initial Lagrangian underdensities whose statistics are
straightforward to compute.  Voids may be detected at high redshift as
a deficit of Lyman-$\alpha$ forest absorption features in QSO spectra.
The Sloan Digital Sky Survey (SDSS) has already measured spectra for
high redshift QSO's over a roughly $\sim$8000 deg$^2$ area,
corresponding to a volume of $\gtrsim 30 ({\rm Gpc}/h)^3$
\cite{sdss_lyaf}.  Each QSO spectrum typically probes $\sim 400
h^{-1}$ Mpc, and the typical transverse separation between QSO
sightlines in SDSS is $\sim 100 h^{-1}$ Mpc, (P.\ McDonald, priv.\
comm.)  so measurements of the clustering of $\sim 10$ Mpc-sized voids
on $\sim$ Gpc scales may already be feasible.

Finally, we note that our conclusions are based on simulations
implementing a very specific type of local primordial nongaussianity
quantified by the $\fnl$ parameter. The validity of our conclusions in the
context of other type of primordial nongaussianity is the subject of
ongoing studies.

\section*{Acknowledgments}
We are grateful to the organisers and participants of the ``Life
beyond the Gaussian'' workshop held at KICP (Chicago, June 2007) where
a preliminary version of this work was presented and discussed. We
thank Xuelei Chen and the National Astronomical Observatories of the
Chinese Academy of Sciences in Beijing where a part of this work was
completed. Finally we thank Niayesh Afshordi, Neil Barnaby, Wayne Hu,
Lam Hui, Nikhil
Padmanabhan, Uros Seljak, Robert Smith, and Emiliano Sefusatti for useful
conversations, Wayne Hu for providing the CMB Fisher matrix, and Pat
McDonald for discussions of the Lyman-$\alpha$ forest.  All
simulations were performed on CITA's Sunnyvale cluster, funded by the
Canada Foundation for Innovation and the Ontario Research Fund for
Research Infrastructure.  This work was supported by the
Canadian Institute for Theoretical Astrophysics (CITA) and the Natural
Sciences and Engineering Research Council of Canada (NSERC).

\appendix*
\section{The abundance and clustering of high peaks}

In this appendix we derive analytic expressions for the abundance 
and clustering of regions above the spherical collapse threshold
$\delta_c$.   We first review previous results derived for Gaussian
statistics, and then show how they are modified by $\fnl$ nongaussianity.

\subsection{Review of Gaussian results} 

We begin by identifying massive halos at late times with high peaks in
the initial density distribution $\delta({\bm x})\equiv\delta\rho({\bm
x})/{\bar\rho}$.  Note that we work entirely in early-time, Lagrangian
coordinates ${\bm x}$ in this section rather than late-time, Eulerian
coordinates.  Earlier work \cite{press-schechter,kaiser84,bbks,bcek}
has shown that the abundance of peaks above threshold 
$\delta_c\approx 1.686$
reasonably describes (at the order-of-magnitude level) the statistics
of halos forming at subsequent times.  We briefly review some of these
previous results, as the methods will be used in our analysis.

Following the \citet{press-schechter} ansatz, we smooth the
density field and assume that density peaks with $\delta>\delta_c$
produce halos.  The density smoothed on scale $R$ is given by
\begin{equation}
\delta_R({\bm x})=(2\pi)^{-3}\int d^3{\bm k}\ \delta_{\bm k}
W(kR) e^{i{\bm k}\cdot{\bm x}},
\end{equation}
where $W(x)$ is some smoothing window, e.g. top-hat or Gaussian.
Assuming that the Fourier modes $\delta_{\bm k}$ are Gaussian
distributed with power spectrum $P(k)=\langle|\delta_k|^2\rangle$,
then $\delta_R({\bm x})$ also has a Gaussian distribution, with
variance
\begin{equation}
\sigma_\delta^2(R)=\langle\delta_R^2\rangle
=\int d\ln k\frac{k^3P(k)}{2\pi^2}W^2(kR).
\end{equation}
The probability $P_1$ for a given randomly selected region to exceed the threshold $\delta_c$
is then simply the integral of the Gaussian probability distribution,
\begin{eqnarray}
P_1&=&\int_{\delta_c}^{\infty} d\delta \frac{dP}{d\delta}
=\int_{\nu_c}^{\infty}d\nu (2\pi)^{-1/2}e^{-\nu^2/2}\nonumber\\
&=&\frac{1}{2}{\rm erfc}\left(\frac{\nu_c}{\sqrt{2}}\right),
\end{eqnarray}
where $\nu=\delta/\sigma_\delta$, and similarly
$\nu_c=\delta_c/\sigma_\delta$. 

The same power spectrum $P(k)$ describing the density variance
$\sigma_\delta^2(R)$ also gives the matter correlation function
$\xi(r_{12})$, and, following an elegant argument by
\citet{kaiser84}, can also be used to determine the correlation
function of rare peaks.  Let us compute the probability $P_2$ that two
randomly selected regions separated by distance $r_{12}\gg R$ are both above threshold.  Again,
this is simply an integral over the (joint) Gaussian distribution:
\begin{equation}
P_2 = \int_{\delta_c}^{\infty}d\delta_1\int_{\delta_c}^{\infty}d\delta_2
\frac{\exp\left(-\frac{1}{2}{\bm\delta}\cdot{\bm\Sigma}^{-1}
\cdot{\bm\delta}\right)}{2\pi|{\bf\Sigma}|^{1/2}}
\end{equation}
where ${\bm\delta}=(\delta_1,\delta_2)$, the covariance matrix is
given by 
\begin{equation}
{\bf\Sigma}=
\left(\begin{array}{cc}
 \sigma_\delta^2 & \xi\\ \xi & \sigma_\delta^2
\end{array}\right),
\end{equation}
and the matter correlation function $\xi(r_{12})$ is given by
\begin{equation}
\xi(r_{12}) = \int d\ln k \frac{k^3P(k)}{2\pi^2}W^2(kR)j_0(kr_{12}).
\end{equation}
Rescaling the $\delta$'s by their variance, this becomes
\begin{equation}
P_2 = \int_{\nu_c}^{\infty}d\nu_1\int_{\nu_c}^{\infty}d\nu_2
\frac{\exp\left(-\frac{1}{2}{\bm\nu}\cdot{\bf S}^{-1}\cdot{\bm\nu}\right)}
{2\pi|{\bf S}|^{1/2}},
\label{p2}
\end{equation}
where
\begin{equation}
{\bf S}=\left(\begin{array}{cc} 1 & \psi\\ \psi & 1\end{array}\right),
\end{equation}
and we follow the notation of BBKS \cite{bbks} in writing the
normalized correlation function as $\psi(r_{12})=\xi(r_{12})/\sigma_\delta^2$;
note that $\psi < 1$.  

To evaluate the integral in Eq.~(\ref{p2}), we change coordinates to
variables that are uncorrelated.  Using the Cholesky decomposition of
the covariance matrix ${\bf S}={\bf C}^{\sf T}{\bf C}$, we write
${\bm\nu}={\bf C}^{\sf T}{\bm y}$ for new variables ${\bm y}$.  Next,
we rotate coordinates ${\bm y}={\bf R}\cdot{\bm x}$ to bring the point
$(\nu_1=\nu_c,\nu_2=\nu_c)$ along the $x_1$ axis.  Then the integral
becomes 
\begin{eqnarray}
P_2&=&\int d^2{\bm x}\frac{\exp\left(-\frac{1}{2}|{\bm x}|^2\right)}{2\pi}
\Theta\left[(1,0)\cdot{\bf C}^{\sf T}\cdot{\bf R}\cdot{\bm x}-\nu_c\right]
\nonumber\\
&&\Theta\left[(0,1)\cdot{\bf C}^{\sf T}\cdot{\bf R}\cdot{\bm x}-\nu_c\right],
\end{eqnarray}
where the Heaviside function $\Theta$ accounts for the two integration
bounds.  Since $\psi<1$, we can write this as
\begin{eqnarray}
P_2&=&\frac{1}{2\pi}\int_{x_c}^\infty dx_1 e^{-x_1^2/2}
\int_{-c (x_1-x_c)}^{c (x_1-x_c)} dx_2 e^{-x_2^2/2} \nonumber\\
&=& \frac{1}{2\pi}\int_{x_c}^\infty dx_1 e^{-x_1^2/2} f(x_1),
\end{eqnarray}
where $c=\sqrt{(1+\psi)/(1-\psi)}$, and $x_c=\nu_c\sqrt{2/(1+\psi)}$.  
We could easily evaluate the integral for $f(x_1)$ in terms of the
error function, but the resulting integral over $x_1$ would then
not be analytic.  However, we can derive an approximate solution in
the limit $\nu_c\gg 1$.  For integrals of the form
\begin{equation}
I=\int_{x_0}^\infty dx\,e^{-x^2/2} f(x),
\end{equation}
we can construct an asymptotic series by repeated partial
integrations :
\begin{equation}
I\approx
e^{-x_0^2/2}\left[\frac{f(x_0)}{x_0}\left(1-\frac{1}{x_0^2}+\ldots\right)
+\frac{f'(x_0)}{x_0^2}+\ldots\right]\;.
\end{equation}
In our case, $f(x_c)=0$ and $f'(x_c)=2c$.  Therefore, in the limit
$\nu_c\gg 1$, we obtain
\begin{eqnarray}
P_2&\approx&\frac{1}{2\pi}e^{-x_c^2/2}\frac{2c}{x_c^2}\nonumber\\
&=&\frac{1}{2\pi}e^{-\nu_c^2/(1+\psi)}\frac{(1+\psi)^{3/2}}{(1-\psi)^{1/2}}
\nu_c^{-2}\;.
\end{eqnarray}
Comparing this expression to the probability for a single peak to be
above threshold then gives the peak-peak correlation function
$\xi_{\rm pk}$, which in the limit $\nu_c\gg 1$, $\psi\ll 1$ becomes
\begin{eqnarray}
1+\xi_{\rm pk}&=&P_2/P_1^2\approx e^{\nu_c^2(1-1/(1+\psi))}\nonumber\\
&\approx&1+\nu_c^2\psi=1+\frac{\nu_c^2}{\sigma_\delta^2}\xi,
\end{eqnarray}
and therefore the (Lagrangian) bias $b_L^2=\xi_{\rm pk}/\xi$ becomes
\begin{equation}
b_L \approx \nu_c^2/\delta_c.
\label{bL}
\end{equation}

\subsection{Nongaussianity}
\label{NG_analytic}

Our discussion so far has merely reviewed previous results for
Gaussian fluctuations; we now turn to nongaussian fluctuations.  As
noted above, we focus on NG of the form
\begin{equation}
\Phi_{\rm NG}=\phi + \fnl (\phi^2-\langle\phi^2\rangle).
\end{equation}
We adopt the approximation of \S\ref{anal} that the heights of rare
peaks are modified by NG as $\delta_{\rm NG}\approx \delta
[1+2\fnl\phi]$, where in this appendix we adopt the notation that
$\phi$ refers to the primordial potential.  At late times, $\phi$
decays as the growth suppression factor $g(a)$.

Let us first consider the one-point distribution of peaks above
threshold, $\delta_{\rm NG}>\delta_c$.  We express this as an integral
over Gaussian variables $\phi$ and $\delta$; the integration bound
then becomes $\delta_{\rm NG}=\delta\,(1+2\fnl\phi)>\delta_c$, and
using the fact that typically $\fnl |\phi| \ll 1$, we have
$\delta > \delta_c (1-2\fnl\phi )$.  The probability for $\delta_{\rm
  NG}$ to exceed threshold then is
\begin{eqnarray}
\!\!\!
P_1 &=& 
\int \!\! d\phi \! \int_{\delta_c (1-2\fnl\phi)}^\infty \!\! d\delta 
\frac{\exp\left(-\frac{1}{2}
(\phi,\delta)\cdot{\bm\Sigma}^{-1}\cdot(\phi,\delta)\right)}
{2\pi|{\bm\Sigma}|^{1/2}} \nonumber\\
&=&\int \!\! d\mu \! \int_{\nu_c-\eta\mu}^\infty \!\!d\nu
\frac{\exp\left(-\frac{1}{2}
(\mu,\nu)\cdot{\bf S}^{-1}\cdot(\mu,\nu)\right)}{2\pi|{\bf
S}|^{1/2}},
\end{eqnarray}
where $\mu=\phi/\sigma_\phi$, $\nu=\delta/\sigma_\delta$, 
$\nu_c=\delta_c/\sigma_\delta$, and $\eta=2\fnl\sigma_\phi\nu_c$.  We
write the off-diagonal part of the normalized covariance matrix 
${\bf S}$ as $\langle\mu\nu\rangle=r$, where the cross-correlation
coefficient $r$ is not to be confused with the peak-peak separation
$r_{12}$ appearing above and below.  By a coordinate
transformation, we can orient the integration bound along a single
axis.  Changing coordinates from $(\mu,\nu)$ to $(\mu,v=\nu+\eta\mu)$, 
and noting that the variance of $v$ is $\langle v^2\rangle=1+2\eta r +
\eta^2$, we can integrate out $\mu$ to obtain
\begin{equation}
P_1=\frac{1}{\sqrt{2\pi}}\int_{x_c}^\infty dx\, e^{-x^2/2} 
=\frac{1}{2}{\rm erfc}\left(\frac{x_c}{\sqrt{2}}\right).
\end{equation}
where $x_c=\nu_c/\sqrt{1+2\eta r+\eta^2}$.
The effect of NG on the peak abundance is therefore simply to rescale
the threshold density by a mass- and redshift-dependent factor.

Next, we turn to the peak-peak correlation function.  As in the
Gaussian case, we write the probability for two points both to be
above threshold as
\begin{eqnarray}
P_2 &=&\int d^4{\bm u}\frac{\exp\left(-\frac{1}{2}
{\bm u}\cdot{\bm\Sigma}\cdot{\bm u}\right)}{(2\pi)^2|{\bm\Sigma}|^{1/2}}
\Theta(\nu_1+\eta\mu_1-\nu_c)
\nonumber\\
&& \times \Theta(\nu_2+\eta\mu_2-\nu_c),
\end{eqnarray}
where ${\bm u}=(\mu_1,\mu_2,\nu_1,\nu_2)$, and the notation is
otherwise the same as above.  We write the off-diagonal parts of the
normalized covariance matrix as $\langle\nu_1\nu_2\rangle=\psi$, 
$\langle\mu_1\mu_2\rangle=\gamma$, $\langle\nu_1\mu_1\rangle=r$, and
$\langle\nu_1\mu_2\rangle=\beta$. As above, we change
variables from $\nu_i$ to $v_i=\nu_i+\eta\mu_i$ to align the
integration bounds along the density coordinate axes.  This allows us
to integrate out the two potential variables, leaving behind a 2-D
integral.  Rescaling the remaining two variables by their (identical)
variance and again writing $x_c=\nu_c/\sqrt{1+2\eta r+\eta^2}$, brings
the integral to the form
\begin{equation}
\int_{x_c}^\infty dx_1 \int_{x_c}^\infty dx_2 
\frac{\exp\left(-\frac{1}{2}{\bm x}\cdot{\bf S}\cdot{\bm x}\right)}{2\pi 
|{\bf S}|^{1/2}}
\label{p2NG}
\end{equation}
where the off-diagonal component of ${\bf S}$ is $\langle x_1
x_2\rangle=\chi$ given by
\begin{equation}
\chi=\frac{\psi+2\eta\beta+\eta^2\gamma}{1+2\eta r + \eta^2}\;.
\end{equation}
The form of equation (\ref{p2NG}) is identical to Eq.~(\ref{p2}),
with $\nu_c\rightarrow x_c$ and $\psi\rightarrow \chi$.  So we can
immediately write down the approximate solution,
\begin{equation}
P_2\approx \frac{1}{2\pi}
e^{-x_c^2/(1+\chi)}x_c^{-2}\frac{(1+\chi)^{3/2}}{(1-\chi)^{1/2}}.
\end{equation}
Comparing with the single-peak probability, we obtain the peak-peak
correlation function, in the limit $\nu_c\gg 1$, $\chi\ll 1$ :
\begin{equation}
1+\xi_{\rm pk} = P_2/P_1^2\approx 1+x_c^2 \chi,
\end{equation}
which to lowest order in $\eta$ becomes
\begin{eqnarray}
\xi_{\rm pk}&\approx&\nu_c^2[\psi + 2\eta(\beta-2r\psi)]\nonumber\\
&\approx&\nu_c^2(\psi + 2\eta\beta)\nonumber\\
&=& b_L^2 (\xi_{\delta\delta} + 4 \fnl \delta_c \xi_{\phi\delta})\;,
\end{eqnarray}
where $b_L$ was the Lagrangian bias obtained for Gaussian peaks,
c.f.\ Eq.~(\ref{bL}).  Note that in going from the first line to the
second, we neglect $r\psi$ relative to $\beta$ since
$\xi_{\delta\delta}(r_{12})/\sigma_{\delta\delta}$ is smaller than 
$\xi_{\phi\delta}(r_{12})/\sigma_{\phi\delta}$ by a factor scaling
like $(R/r_{12})^2$, where $R$ is the smoothing scale of the peak and
$r_{12}$ is the peak-peak separation.  

The peak-peak correlation function is now no 
longer simply proportional to the matter correlation function,
implying that the peak bias is not independent of scale.  Fourier
transforming this expression gives the peak power spectrum,
\begin{equation}
P_{\rm pk}=b_L^2 (P_{\delta\delta} + 4 \fnl \delta_c P_{\phi\delta})
\end{equation}
which gives a scale-dependent change in the bias due to NG of
\begin{eqnarray}
\Delta b(k) &=& 2 b_L \fnl\delta_c \frac{P_{\phi\delta}}{P_{\delta\delta}}\nonumber\\
&=& 2 b_L \fnl\delta_c \frac{3\Omega_m}{2ag\,r_H^2k^2}\;,
\label{bk}
\end{eqnarray}
where we have used the relation between the potential-density
cross-spectrum and the matter power spectrum
$P_{\phi\delta}=(3\Omega_m/2ag\,r_H^2k^2)P_{\delta\delta}$,
arising from the Poisson equation. 
The total Lagrangian bias is then $b_L(k)=b_L + \Delta b(k)$.


\begin{thebibliography}{114}
\expandafter\ifx\csname natexlab\endcsname\relax\def\natexlab#1{#1}\fi
\expandafter\ifx\csname bibnamefont\endcsname\relax
  \def\bibnamefont#1{#1}\fi
\expandafter\ifx\csname bibfnamefont\endcsname\relax
  \def\bibfnamefont#1{#1}\fi
\expandafter\ifx\csname citenamefont\endcsname\relax
  \def\citenamefont#1{#1}\fi
\expandafter\ifx\csname url\endcsname\relax
  \def\url#1{\texttt{#1}}\fi
\expandafter\ifx\csname urlprefix\endcsname\relax\def\urlprefix{URL }\fi
\providecommand{\bibinfo}[2]{#2}
\providecommand{\eprint}[2][]{\url{#2}}

\bibitem[{\citenamefont{{Maldacena}}(2003)}]{maldacena}
\bibinfo{author}{\bibfnamefont{J.}~\bibnamefont{{Maldacena}}},
  \bibinfo{journal}{Journal of High Energy Physics}
  \textbf{\bibinfo{volume}{5}}, \bibinfo{pages}{13} (\bibinfo{year}{2003}),
  \eprint{arXiv:astro-ph/0210603}.

\bibitem[{\citenamefont{Acquaviva et~al.}(2003)\citenamefont{Acquaviva,
  Bartolo, Matarrese, and Riotto}}]{Acquaviva:2002ud}
\bibinfo{author}{\bibfnamefont{V.}~\bibnamefont{Acquaviva}},
  \bibinfo{author}{\bibfnamefont{N.}~\bibnamefont{Bartolo}},
  \bibinfo{author}{\bibfnamefont{S.}~\bibnamefont{Matarrese}},
  \bibnamefont{and} \bibinfo{author}{\bibfnamefont{A.}~\bibnamefont{Riotto}},
  \bibinfo{journal}{Nucl. Phys.} \textbf{\bibinfo{volume}{B667}},
  \bibinfo{pages}{119} (\bibinfo{year}{2003}), \eprint{astro-ph/0209156}.

\bibitem[{\citenamefont{Creminelli}(2003)}]{Creminelli:2003iq}
\bibinfo{author}{\bibfnamefont{P.}~\bibnamefont{Creminelli}},
  \bibinfo{journal}{JCAP} \textbf{\bibinfo{volume}{0310}}, \bibinfo{pages}{003}
  (\bibinfo{year}{2003}), \eprint{astro-ph/0306122}.

\bibitem[{\citenamefont{Lyth and
  Rodriguez}(2005{\natexlab{a}})}]{Lyth_Rodriguez}
\bibinfo{author}{\bibfnamefont{D.~H.} \bibnamefont{Lyth}} \bibnamefont{and}
  \bibinfo{author}{\bibfnamefont{Y.}~\bibnamefont{Rodriguez}},
  \bibinfo{journal}{Phys. Rev. Lett.} \textbf{\bibinfo{volume}{95}},
  \bibinfo{pages}{121302} (\bibinfo{year}{2005}{\natexlab{a}}),
  \eprint{astro-ph/0504045}.

\bibitem[{\citenamefont{Seery and Lidsey}(2005)}]{Seery_Lidsey}
\bibinfo{author}{\bibfnamefont{D.}~\bibnamefont{Seery}} \bibnamefont{and}
  \bibinfo{author}{\bibfnamefont{J.~E.} \bibnamefont{Lidsey}},
  \bibinfo{journal}{JCAP} \textbf{\bibinfo{volume}{0506}}, \bibinfo{pages}{003}
  (\bibinfo{year}{2005}), \eprint{astro-ph/0503692}.

\bibitem[{\citenamefont{{Spergel} et~al.}(2007)\citenamefont{{Spergel}, {Bean},
  {Dor{\'e}}, {Nolta}, {Bennett}, {Dunkley}, {Hinshaw}, {Jarosik}, {Komatsu},
  {Page} et~al.}}]{wmap3}
\bibinfo{author}{\bibfnamefont{D.~N.} \bibnamefont{{Spergel}}},
  \bibinfo{author}{\bibfnamefont{R.}~\bibnamefont{{Bean}}},
  \bibinfo{author}{\bibfnamefont{O.}~\bibnamefont{{Dor{\'e}}}},
  \bibinfo{author}{\bibfnamefont{M.~R.} \bibnamefont{{Nolta}}},
  \bibinfo{author}{\bibfnamefont{C.~L.} \bibnamefont{{Bennett}}},
  \bibinfo{author}{\bibfnamefont{J.}~\bibnamefont{{Dunkley}}},
  \bibinfo{author}{\bibfnamefont{G.}~\bibnamefont{{Hinshaw}}},
  \bibinfo{author}{\bibfnamefont{N.}~\bibnamefont{{Jarosik}}},
  \bibinfo{author}{\bibfnamefont{E.}~\bibnamefont{{Komatsu}}},
  \bibinfo{author}{\bibfnamefont{L.}~\bibnamefont{{Page}}},
  \bibnamefont{et~al.}, \bibinfo{journal}{\apjs}
  \textbf{\bibinfo{volume}{170}}, \bibinfo{pages}{377} (\bibinfo{year}{2007}),
  \eprint{arXiv:astro-ph/0603449}.

\bibitem[{\citenamefont{Creminelli
  et~al.}(2007{\natexlab{a}})\citenamefont{Creminelli, Senatore, Zaldarriaga,
  and Tegmark}}]{Creminelli_wmap}
\bibinfo{author}{\bibfnamefont{P.}~\bibnamefont{Creminelli}},
  \bibinfo{author}{\bibfnamefont{L.}~\bibnamefont{Senatore}},
  \bibinfo{author}{\bibfnamefont{M.}~\bibnamefont{Zaldarriaga}},
  \bibnamefont{and} \bibinfo{author}{\bibfnamefont{M.}~\bibnamefont{Tegmark}},
  \bibinfo{journal}{JCAP} \textbf{\bibinfo{volume}{0703}}, \bibinfo{pages}{005}
  (\bibinfo{year}{2007}{\natexlab{a}}), \eprint{astro-ph/0610600}.

\bibitem[{\citenamefont{Arkani-Hamed et~al.}(2004)\citenamefont{Arkani-Hamed,
  Creminelli, Mukohyama, and Zaldarriaga}}]{ArkaniHamed:2003uz}
\bibinfo{author}{\bibfnamefont{N.}~\bibnamefont{Arkani-Hamed}},
  \bibinfo{author}{\bibfnamefont{P.}~\bibnamefont{Creminelli}},
  \bibinfo{author}{\bibfnamefont{S.}~\bibnamefont{Mukohyama}},
  \bibnamefont{and}
  \bibinfo{author}{\bibfnamefont{M.}~\bibnamefont{Zaldarriaga}},
  \bibinfo{journal}{JCAP} \textbf{\bibinfo{volume}{0404}}, \bibinfo{pages}{001}
  (\bibinfo{year}{2004}), \eprint{hep-th/0312100}.

\bibitem[{\citenamefont{Bartolo
  et~al.}(2004{\natexlab{a}})\citenamefont{Bartolo, Matarrese, and
  Riotto}}]{Bartolo:2003jx}
\bibinfo{author}{\bibfnamefont{N.}~\bibnamefont{Bartolo}},
  \bibinfo{author}{\bibfnamefont{S.}~\bibnamefont{Matarrese}},
  \bibnamefont{and} \bibinfo{author}{\bibfnamefont{A.}~\bibnamefont{Riotto}},
  \bibinfo{journal}{Phys. Rev.} \textbf{\bibinfo{volume}{D69}},
  \bibinfo{pages}{043503} (\bibinfo{year}{2004}{\natexlab{a}}),
  \eprint{hep-ph/0309033}.

\bibitem[{\citenamefont{Lyth and Rodriguez}(2005{\natexlab{b}})}]{Lyth:2005du}
\bibinfo{author}{\bibfnamefont{D.~H.} \bibnamefont{Lyth}} \bibnamefont{and}
  \bibinfo{author}{\bibfnamefont{Y.}~\bibnamefont{Rodriguez}},
  \bibinfo{journal}{Phys. Rev.} \textbf{\bibinfo{volume}{D71}},
  \bibinfo{pages}{123508} (\bibinfo{year}{2005}{\natexlab{b}}),
  \eprint{astro-ph/0502578}.

\bibitem[{\citenamefont{Rigopoulos et~al.}(2006)\citenamefont{Rigopoulos,
  Shellard, and van Tent}}]{Rigopoulos:2005ae}
\bibinfo{author}{\bibfnamefont{G.~I.} \bibnamefont{Rigopoulos}},
  \bibinfo{author}{\bibfnamefont{E.~P.~S.} \bibnamefont{Shellard}},
  \bibnamefont{and} \bibinfo{author}{\bibfnamefont{B.~J.~W.} \bibnamefont{van
  Tent}}, \bibinfo{journal}{Phys. Rev.} \textbf{\bibinfo{volume}{D73}},
  \bibinfo{pages}{083522} (\bibinfo{year}{2006}), \eprint{astro-ph/0506704}.

\bibitem[{\citenamefont{Allen et~al.}(2006)\citenamefont{Allen, Gupta, and
  Wands}}]{Allen:2005ye}
\bibinfo{author}{\bibfnamefont{L.~E.} \bibnamefont{Allen}},
  \bibinfo{author}{\bibfnamefont{S.}~\bibnamefont{Gupta}}, \bibnamefont{and}
  \bibinfo{author}{\bibfnamefont{D.}~\bibnamefont{Wands}},
  \bibinfo{journal}{JCAP} \textbf{\bibinfo{volume}{0601}}, \bibinfo{pages}{006}
  (\bibinfo{year}{2006}), \eprint{astro-ph/0509719}.

\bibitem[{\citenamefont{Chen}(2005)}]{Chen_DBI}
\bibinfo{author}{\bibfnamefont{X.}~\bibnamefont{Chen}}, \bibinfo{journal}{Phys.
  Rev.} \textbf{\bibinfo{volume}{D72}}, \bibinfo{pages}{123518}
  (\bibinfo{year}{2005}), \eprint{astro-ph/0507053}.

\bibitem[{\citenamefont{Barnaby and Cline}(2006)}]{Barnaby:2006cq}
\bibinfo{author}{\bibfnamefont{N.}~\bibnamefont{Barnaby}} \bibnamefont{and}
  \bibinfo{author}{\bibfnamefont{J.~M.} \bibnamefont{Cline}},
  \bibinfo{journal}{Phys. Rev.} \textbf{\bibinfo{volume}{D73}},
  \bibinfo{pages}{106012} (\bibinfo{year}{2006}), \eprint{astro-ph/0601481}.

\bibitem[{\citenamefont{Barnaby and
  Cline}(2007{\natexlab{a}})}]{Barnaby:2006km}
\bibinfo{author}{\bibfnamefont{N.}~\bibnamefont{Barnaby}} \bibnamefont{and}
  \bibinfo{author}{\bibfnamefont{J.~M.} \bibnamefont{Cline}},
  \bibinfo{journal}{Phys. Rev.} \textbf{\bibinfo{volume}{D75}},
  \bibinfo{pages}{086004} (\bibinfo{year}{2007}{\natexlab{a}}),
  \eprint{astro-ph/0611750}.

\bibitem[{\citenamefont{Barnaby and
  Cline}(2007{\natexlab{b}})}]{Barnaby:2007yb}
\bibinfo{author}{\bibfnamefont{N.}~\bibnamefont{Barnaby}} \bibnamefont{and}
  \bibinfo{author}{\bibfnamefont{J.~M.} \bibnamefont{Cline}}
  (\bibinfo{year}{2007}{\natexlab{b}}), \eprint{arXiv:0704.3426 [hep-th]}.

\bibitem[{\citenamefont{Sasaki et~al.}(2006)\citenamefont{Sasaki, Valiviita,
  and Wands}}]{Sasaki:2006kq}
\bibinfo{author}{\bibfnamefont{M.}~\bibnamefont{Sasaki}},
  \bibinfo{author}{\bibfnamefont{J.}~\bibnamefont{Valiviita}},
  \bibnamefont{and} \bibinfo{author}{\bibfnamefont{D.}~\bibnamefont{Wands}},
  \bibinfo{journal}{Phys. Rev.} \textbf{\bibinfo{volume}{D74}},
  \bibinfo{pages}{103003} (\bibinfo{year}{2006}), \eprint{astro-ph/0607627}.

\bibitem[{\citenamefont{Chen et~al.}(2007{\natexlab{a}})\citenamefont{Chen,
  Huang, Kachru, and Shiu}}]{Chen:2006nt}
\bibinfo{author}{\bibfnamefont{X.}~\bibnamefont{Chen}},
  \bibinfo{author}{\bibfnamefont{M.-x.} \bibnamefont{Huang}},
  \bibinfo{author}{\bibfnamefont{S.}~\bibnamefont{Kachru}}, \bibnamefont{and}
  \bibinfo{author}{\bibfnamefont{G.}~\bibnamefont{Shiu}},
  \bibinfo{journal}{JCAP} \textbf{\bibinfo{volume}{0701}}, \bibinfo{pages}{002}
  (\bibinfo{year}{2007}{\natexlab{a}}), \eprint{hep-th/0605045}.

\bibitem[{\citenamefont{Chen et~al.}(2007{\natexlab{b}})\citenamefont{Chen,
  Easther, and Lim}}]{Chen_Easther_Lim}
\bibinfo{author}{\bibfnamefont{X.}~\bibnamefont{Chen}},
  \bibinfo{author}{\bibfnamefont{R.}~\bibnamefont{Easther}}, \bibnamefont{and}
  \bibinfo{author}{\bibfnamefont{E.~A.} \bibnamefont{Lim}},
  \bibinfo{journal}{JCAP} \textbf{\bibinfo{volume}{0706}}, \bibinfo{pages}{023}
  (\bibinfo{year}{2007}{\natexlab{b}}), \eprint{astro-ph/0611645}.

\bibitem[{\citenamefont{Battefeld and Easther}(2007)}]{Battefeld_Easther}
\bibinfo{author}{\bibfnamefont{T.}~\bibnamefont{Battefeld}} \bibnamefont{and}
  \bibinfo{author}{\bibfnamefont{R.}~\bibnamefont{Easther}},
  \bibinfo{journal}{JCAP} \textbf{\bibinfo{volume}{0703}}, \bibinfo{pages}{020}
  (\bibinfo{year}{2007}), \eprint{astro-ph/0610296}.

\bibitem[{\citenamefont{Assadullahi et~al.}(2007)\citenamefont{Assadullahi,
  Valiviita, and Wands}}]{Assadullahi:2007uw}
\bibinfo{author}{\bibfnamefont{H.}~\bibnamefont{Assadullahi}},
  \bibinfo{author}{\bibfnamefont{J.}~\bibnamefont{Valiviita}},
  \bibnamefont{and} \bibinfo{author}{\bibfnamefont{D.}~\bibnamefont{Wands}}
  (\bibinfo{year}{2007}), \eprint{arXiv:0708.0223 [hep-ph]}.

\bibitem[{\citenamefont{Battefeld and Battefeld}(2007)}]{Battefeld:2007en}
\bibinfo{author}{\bibfnamefont{D.}~\bibnamefont{Battefeld}} \bibnamefont{and}
  \bibinfo{author}{\bibfnamefont{T.}~\bibnamefont{Battefeld}},
  \bibinfo{journal}{JCAP} \textbf{\bibinfo{volume}{0705}}, \bibinfo{pages}{012}
  (\bibinfo{year}{2007}), \eprint{hep-th/0703012}.

\bibitem[{\citenamefont{Bean et~al.}(2007)\citenamefont{Bean, Shandera,
  Henry~Tye, and Xu}}]{Shandera}
\bibinfo{author}{\bibfnamefont{R.}~\bibnamefont{Bean}},
  \bibinfo{author}{\bibfnamefont{S.~E.} \bibnamefont{Shandera}},
  \bibinfo{author}{\bibfnamefont{S.~H.} \bibnamefont{Henry~Tye}},
  \bibnamefont{and} \bibinfo{author}{\bibfnamefont{J.}~\bibnamefont{Xu}},
  \bibinfo{journal}{JCAP} \textbf{\bibinfo{volume}{0705}}, \bibinfo{pages}{004}
  (\bibinfo{year}{2007}), \eprint{hep-th/0702107}.

\bibitem[{\citenamefont{Bartolo
  et~al.}(2004{\natexlab{b}})\citenamefont{Bartolo, Komatsu, Matarrese, and
  Riotto}}]{Bartolo:2004if}
\bibinfo{author}{\bibfnamefont{N.}~\bibnamefont{Bartolo}},
  \bibinfo{author}{\bibfnamefont{E.}~\bibnamefont{Komatsu}},
  \bibinfo{author}{\bibfnamefont{S.}~\bibnamefont{Matarrese}},
  \bibnamefont{and} \bibinfo{author}{\bibfnamefont{A.}~\bibnamefont{Riotto}},
  \bibinfo{journal}{Phys. Rept.} \textbf{\bibinfo{volume}{402}},
  \bibinfo{pages}{103} (\bibinfo{year}{2004}{\natexlab{b}}),
  \eprint{astro-ph/0406398}.

\bibitem[{\citenamefont{Babich}(2005)}]{Babich:2005en}
\bibinfo{author}{\bibfnamefont{D.}~\bibnamefont{Babich}},
  \bibinfo{journal}{Phys. Rev.} \textbf{\bibinfo{volume}{D72}},
  \bibinfo{pages}{043003} (\bibinfo{year}{2005}), \eprint{astro-ph/0503375}.

\bibitem[{\citenamefont{Babich et~al.}(2004)\citenamefont{Babich, Creminelli,
  and Zaldarriaga}}]{Babich_shape}
\bibinfo{author}{\bibfnamefont{D.}~\bibnamefont{Babich}},
  \bibinfo{author}{\bibfnamefont{P.}~\bibnamefont{Creminelli}},
  \bibnamefont{and}
  \bibinfo{author}{\bibfnamefont{M.}~\bibnamefont{Zaldarriaga}},
  \bibinfo{journal}{JCAP} \textbf{\bibinfo{volume}{0408}}, \bibinfo{pages}{009}
  (\bibinfo{year}{2004}), \eprint{astro-ph/0405356}.

\bibitem[{\citenamefont{Creminelli
  et~al.}(2007{\natexlab{b}})\citenamefont{Creminelli, Senatore, and
  Zaldarriaga}}]{Creminelli_estimators}
\bibinfo{author}{\bibfnamefont{P.}~\bibnamefont{Creminelli}},
  \bibinfo{author}{\bibfnamefont{L.}~\bibnamefont{Senatore}}, \bibnamefont{and}
  \bibinfo{author}{\bibfnamefont{M.}~\bibnamefont{Zaldarriaga}},
  \bibinfo{journal}{JCAP} \textbf{\bibinfo{volume}{0703}}, \bibinfo{pages}{019}
  (\bibinfo{year}{2007}{\natexlab{b}}), \eprint{astro-ph/0606001}.

\bibitem[{\citenamefont{Smith and
  Zaldarriaga}(2006{\natexlab{a}})}]{Smith_Zaldarriaga}
\bibinfo{author}{\bibfnamefont{K.~M.} \bibnamefont{Smith}} \bibnamefont{and}
  \bibinfo{author}{\bibfnamefont{M.}~\bibnamefont{Zaldarriaga}}
  (\bibinfo{year}{2006}{\natexlab{a}}), \eprint{astro-ph/0612571}.

\bibitem[{\citenamefont{Fergusson and Shellard}(2006)}]{Fergusson_Shellard}
\bibinfo{author}{\bibfnamefont{J.~R.} \bibnamefont{Fergusson}}
  \bibnamefont{and} \bibinfo{author}{\bibfnamefont{E.~P.~S.}
  \bibnamefont{Shellard}} (\bibinfo{year}{2006}), \eprint{astro-ph/0612713}.

\bibitem[{\citenamefont{Falk et~al.}(1993)\citenamefont{Falk, Rangarajan, and
  Srednicki}}]{Falk_Ran_Sre}
\bibinfo{author}{\bibfnamefont{T.}~\bibnamefont{Falk}},
  \bibinfo{author}{\bibfnamefont{R.}~\bibnamefont{Rangarajan}},
  \bibnamefont{and}
  \bibinfo{author}{\bibfnamefont{M.}~\bibnamefont{Srednicki}},
  \bibinfo{journal}{Astrophys. J.} \textbf{\bibinfo{volume}{403}},
  \bibinfo{pages}{L1} (\bibinfo{year}{1993}), \eprint{astro-ph/9208001}.

\bibitem[{\citenamefont{Luo and Schramm}(1993)}]{Luo_Schramm}
\bibinfo{author}{\bibfnamefont{X.-c.} \bibnamefont{Luo}} \bibnamefont{and}
  \bibinfo{author}{\bibfnamefont{D.~N.} \bibnamefont{Schramm}},
  \bibinfo{journal}{Phys. Rev. Lett.} \textbf{\bibinfo{volume}{71}},
  \bibinfo{pages}{1124} (\bibinfo{year}{1993}), \eprint{astro-ph/9305009}.

\bibitem[{\citenamefont{Gangui et~al.}(1994)\citenamefont{Gangui, Lucchin,
  Matarrese, and Mollerach}}]{Gangui_etal}
\bibinfo{author}{\bibfnamefont{A.}~\bibnamefont{Gangui}},
  \bibinfo{author}{\bibfnamefont{F.}~\bibnamefont{Lucchin}},
  \bibinfo{author}{\bibfnamefont{S.}~\bibnamefont{Matarrese}},
  \bibnamefont{and}
  \bibinfo{author}{\bibfnamefont{S.}~\bibnamefont{Mollerach}},
  \bibinfo{journal}{Astrophys. J.} \textbf{\bibinfo{volume}{430}},
  \bibinfo{pages}{447} (\bibinfo{year}{1994}), \eprint{astro-ph/9312033}.

\bibitem[{\citenamefont{Wang and Kamionkowski}(2000)}]{Wang_Kam}
\bibinfo{author}{\bibfnamefont{L.-M.} \bibnamefont{Wang}} \bibnamefont{and}
  \bibinfo{author}{\bibfnamefont{M.}~\bibnamefont{Kamionkowski}},
  \bibinfo{journal}{Phys. Rev.} \textbf{\bibinfo{volume}{D61}},
  \bibinfo{pages}{063504} (\bibinfo{year}{2000}), \eprint{astro-ph/9907431}.

\bibitem[{\citenamefont{Verde et~al.}(2001)\citenamefont{Verde, Jimenez,
  Kamionkowski, and Matarrese}}]{Verde:2000vr}
\bibinfo{author}{\bibfnamefont{L.}~\bibnamefont{Verde}},
  \bibinfo{author}{\bibfnamefont{R.}~\bibnamefont{Jimenez}},
  \bibinfo{author}{\bibfnamefont{M.}~\bibnamefont{Kamionkowski}},
  \bibnamefont{and}
  \bibinfo{author}{\bibfnamefont{S.}~\bibnamefont{Matarrese}},
  \bibinfo{journal}{Mon. Not. Roy. Astron. Soc.}
  \textbf{\bibinfo{volume}{325}}, \bibinfo{pages}{412} (\bibinfo{year}{2001}),
  \eprint{astro-ph/0011180}.

\bibitem[{\citenamefont{Scoccimarro et~al.}(2004)\citenamefont{Scoccimarro,
  Sefusatti, and Zaldarriaga}}]{Scoccimarro:2003wn}
\bibinfo{author}{\bibfnamefont{R.}~\bibnamefont{Scoccimarro}},
  \bibinfo{author}{\bibfnamefont{E.}~\bibnamefont{Sefusatti}},
  \bibnamefont{and}
  \bibinfo{author}{\bibfnamefont{M.}~\bibnamefont{Zaldarriaga}},
  \bibinfo{journal}{Phys. Rev.} \textbf{\bibinfo{volume}{D69}},
  \bibinfo{pages}{103513} (\bibinfo{year}{2004}), \eprint{astro-ph/0312286}.

\bibitem[{\citenamefont{Sefusatti and Komatsu}(2007)}]{Sefusatti:2007ih}
\bibinfo{author}{\bibfnamefont{E.}~\bibnamefont{Sefusatti}} \bibnamefont{and}
  \bibinfo{author}{\bibfnamefont{E.}~\bibnamefont{Komatsu}}
  (\bibinfo{year}{2007}), \eprint{arXiv:0705.0343 [astro-ph]}.

\bibitem[{\citenamefont{Lucchin and Matarrese}(1988)}]{Lucchin:1987yv}
\bibinfo{author}{\bibfnamefont{F.}~\bibnamefont{Lucchin}} \bibnamefont{and}
  \bibinfo{author}{\bibfnamefont{S.}~\bibnamefont{Matarrese}},
  \bibinfo{journal}{Astrophys. J.} \textbf{\bibinfo{volume}{330}},
  \bibinfo{pages}{535} (\bibinfo{year}{1988}).

\bibitem[{\citenamefont{Robinson and Baker}(1999)}]{Robinson:1999se}
\bibinfo{author}{\bibfnamefont{J.}~\bibnamefont{Robinson}} \bibnamefont{and}
  \bibinfo{author}{\bibfnamefont{J.~E.} \bibnamefont{Baker}}
  (\bibinfo{year}{1999}), \eprint{astro-ph/9905098}.

\bibitem[{\citenamefont{Benson et~al.}(2002)\citenamefont{Benson, Reichardt,
  and Kamionkowski}}]{Benson:2001hc}
\bibinfo{author}{\bibfnamefont{A.~J.} \bibnamefont{Benson}},
  \bibinfo{author}{\bibfnamefont{C.}~\bibnamefont{Reichardt}},
  \bibnamefont{and}
  \bibinfo{author}{\bibfnamefont{M.}~\bibnamefont{Kamionkowski}},
  \bibinfo{journal}{Mon. Not. Roy. Astron. Soc.}
  \textbf{\bibinfo{volume}{331}}, \bibinfo{pages}{71} (\bibinfo{year}{2002}),
  \eprint{astro-ph/0110299}.

\bibitem[{\citenamefont{Matarrese et~al.}(2000)\citenamefont{Matarrese, Verde,
  and Jimenez}}]{Matarrese:2000iz}
\bibinfo{author}{\bibfnamefont{S.}~\bibnamefont{Matarrese}},
  \bibinfo{author}{\bibfnamefont{L.}~\bibnamefont{Verde}}, \bibnamefont{and}
  \bibinfo{author}{\bibfnamefont{R.}~\bibnamefont{Jimenez}},
  \bibinfo{journal}{Astrophys. J.} \textbf{\bibinfo{volume}{541}},
  \bibinfo{pages}{10} (\bibinfo{year}{2000}), \eprint{astro-ph/0001366}.

\bibitem[{\citenamefont{Komatsu et~al.}(2003)}]{Komatsu:2003fd}
\bibinfo{author}{\bibfnamefont{E.}~\bibnamefont{Komatsu}} \bibnamefont{et~al.}
  (\bibinfo{collaboration}{WMAP}), \bibinfo{journal}{Astrophys. J. Suppl.}
  \textbf{\bibinfo{volume}{148}}, \bibinfo{pages}{119} (\bibinfo{year}{2003}),
  \eprint{astro-ph/0302223}.

\bibitem[{\citenamefont{{Verde} et~al.}(2001)\citenamefont{{Verde},
  {Kamionkowski}, {Mohr}, and {Benson}}}]{verde01}
\bibinfo{author}{\bibfnamefont{L.}~\bibnamefont{{Verde}}},
  \bibinfo{author}{\bibfnamefont{M.}~\bibnamefont{{Kamionkowski}}},
  \bibinfo{author}{\bibfnamefont{J.~J.} \bibnamefont{{Mohr}}},
  \bibnamefont{and} \bibinfo{author}{\bibfnamefont{A.~J.}
  \bibnamefont{{Benson}}}, \bibinfo{journal}{\mnras}
  \textbf{\bibinfo{volume}{321}}, \bibinfo{pages}{L7} (\bibinfo{year}{2001}),
  \eprint{arXiv:astro-ph/0007426}.

\bibitem[{\citenamefont{{Rozo} et~al.}(2007)\citenamefont{{Rozo}, {Wechsler},
  {Koester}, {McKay}, {Evrard}, {Johnston}, {Sheldon}, {Annis}, and
  {Frieman}}}]{maxBCG}
\bibinfo{author}{\bibfnamefont{E.}~\bibnamefont{{Rozo}}},
  \bibinfo{author}{\bibfnamefont{R.~H.} \bibnamefont{{Wechsler}}},
  \bibinfo{author}{\bibfnamefont{B.~P.} \bibnamefont{{Koester}}},
  \bibinfo{author}{\bibfnamefont{T.~A.} \bibnamefont{{McKay}}},
  \bibinfo{author}{\bibfnamefont{A.~E.} \bibnamefont{{Evrard}}},
  \bibinfo{author}{\bibfnamefont{D.}~\bibnamefont{{Johnston}}},
  \bibinfo{author}{\bibfnamefont{E.~S.} \bibnamefont{{Sheldon}}},
  \bibinfo{author}{\bibfnamefont{J.}~\bibnamefont{{Annis}}}, \bibnamefont{and}
  \bibinfo{author}{\bibfnamefont{J.~A.} \bibnamefont{{Frieman}}},
  \bibinfo{journal}{ArXiv Astrophysics e-prints}  (\bibinfo{year}{2007}),
  \eprint{astro-ph/0703571}.

\bibitem[{\citenamefont{Koester et~al.}(2007)}]{Koester:2007bg}
\bibinfo{author}{\bibfnamefont{B.}~\bibnamefont{Koester}} \bibnamefont{et~al.}
  (\bibinfo{collaboration}{SDSS}), \bibinfo{journal}{Astrophys. J.}
  \textbf{\bibinfo{volume}{660}}, \bibinfo{pages}{239} (\bibinfo{year}{2007}),
  \eprint{astro-ph/0701265}.

\bibitem[{\citenamefont{Eke et~al.}(2004)}]{Eke:2004ve}
\bibinfo{author}{\bibfnamefont{V.~R.} \bibnamefont{Eke}} \bibnamefont{et~al.}
  (\bibinfo{collaboration}{The 2dFGRS Team}), \bibinfo{journal}{Mon. Not. Roy.
  Astron. Soc.} \textbf{\bibinfo{volume}{348}}, \bibinfo{pages}{866}
  (\bibinfo{year}{2004}), \eprint{astro-ph/0402567}.

\bibitem[{\citenamefont{Yee et~al.}(2007)}]{Yee:2007if}
\bibinfo{author}{\bibfnamefont{H.~K.~C.} \bibnamefont{Yee}}
  \bibnamefont{et~al.} (\bibinfo{collaboration}{RCS-2}) (\bibinfo{year}{2007}),
  \eprint{astro-ph/0701839}.

\bibitem[{\citenamefont{Willis et~al.}(2005)}]{Willis:2005ag}
\bibinfo{author}{\bibfnamefont{J.~P.} \bibnamefont{Willis}}
  \bibnamefont{et~al.}, \bibinfo{journal}{Mon. Not. Roy. Astron. Soc.}
  \textbf{\bibinfo{volume}{363}}, \bibinfo{pages}{675} (\bibinfo{year}{2005}),
  \eprint{astro-ph/0508003}.

\bibitem[{\citenamefont{Valtchanov et~al.}(2004)}]{Valtchanov:2003it}
\bibinfo{author}{\bibfnamefont{I.}~\bibnamefont{Valtchanov}}
  \bibnamefont{et~al.}, \bibinfo{journal}{Astron. Astrophys.}
  \textbf{\bibinfo{volume}{423}}, \bibinfo{pages}{75} (\bibinfo{year}{2004}),
  \eprint{astro-ph/0305192}.

\bibitem[{\citenamefont{{Haiman} et~al.}(2001)\citenamefont{{Haiman}, {Mohr},
  and {Holder}}}]{Haiman_Mohr_Holder}
\bibinfo{author}{\bibfnamefont{Z.}~\bibnamefont{{Haiman}}},
  \bibinfo{author}{\bibfnamefont{J.~J.} \bibnamefont{{Mohr}}},
  \bibnamefont{and} \bibinfo{author}{\bibfnamefont{G.~P.}
  \bibnamefont{{Holder}}}, \bibinfo{journal}{\apj}
  \textbf{\bibinfo{volume}{553}}, \bibinfo{pages}{545} (\bibinfo{year}{2001}),
  \eprint{arXiv:astro-ph/0002336}.

\bibitem[{\citenamefont{Majumdar and Mohr}(2003)}]{Majumdar_Mohr}
\bibinfo{author}{\bibfnamefont{S.}~\bibnamefont{Majumdar}} \bibnamefont{and}
  \bibinfo{author}{\bibfnamefont{J.~J.} \bibnamefont{Mohr}},
  \bibinfo{journal}{Astrophys. J.} \textbf{\bibinfo{volume}{585}},
  \bibinfo{pages}{603} (\bibinfo{year}{2003}), \eprint{astro-ph/0208002}.

\bibitem[{\citenamefont{Wang et~al.}(2004)\citenamefont{Wang, Khoury, Haiman,
  and May}}]{Wang_Haiman}
\bibinfo{author}{\bibfnamefont{S.}~\bibnamefont{Wang}},
  \bibinfo{author}{\bibfnamefont{J.}~\bibnamefont{Khoury}},
  \bibinfo{author}{\bibfnamefont{Z.}~\bibnamefont{Haiman}}, \bibnamefont{and}
  \bibinfo{author}{\bibfnamefont{M.}~\bibnamefont{May}},
  \bibinfo{journal}{Phys. Rev.} \textbf{\bibinfo{volume}{D70}},
  \bibinfo{pages}{123008} (\bibinfo{year}{2004}), \eprint{astro-ph/0406331}.

\bibitem[{\citenamefont{Battye and Weller}(2005)}]{Battye_Weller}
\bibinfo{author}{\bibfnamefont{R.~A.} \bibnamefont{Battye}} \bibnamefont{and}
  \bibinfo{author}{\bibfnamefont{J.}~\bibnamefont{Weller}},
  \bibinfo{journal}{Mon. Not. Roy. Astron. Soc.}
  \textbf{\bibinfo{volume}{362}}, \bibinfo{pages}{171} (\bibinfo{year}{2005}),
  \eprint{astro-ph/0410392}.

\bibitem[{\citenamefont{Lima and Hu}(2005)}]{Lima_Hu_05}
\bibinfo{author}{\bibfnamefont{M.}~\bibnamefont{Lima}} \bibnamefont{and}
  \bibinfo{author}{\bibfnamefont{W.}~\bibnamefont{Hu}}, \bibinfo{journal}{Phys.
  Rev.} \textbf{\bibinfo{volume}{D72}}, \bibinfo{pages}{043006}
  (\bibinfo{year}{2005}), \eprint{astro-ph/0503363}.

\bibitem[{\citenamefont{Marian and Bernstein}(2006)}]{Marian_Bernstein}
\bibinfo{author}{\bibfnamefont{L.}~\bibnamefont{Marian}} \bibnamefont{and}
  \bibinfo{author}{\bibfnamefont{G.~M.} \bibnamefont{Bernstein}},
  \bibinfo{journal}{Phys. Rev.} \textbf{\bibinfo{volume}{D73}},
  \bibinfo{pages}{123525} (\bibinfo{year}{2006}), \eprint{astro-ph/0605746}.

\bibitem[{\citenamefont{Takada and Bridle}(2007)}]{Takada_Bridle}
\bibinfo{author}{\bibfnamefont{M.}~\bibnamefont{Takada}} \bibnamefont{and}
  \bibinfo{author}{\bibfnamefont{S.}~\bibnamefont{Bridle}}
  (\bibinfo{year}{2007}), \eprint{arXiv:0705.0163 [astro-ph]}.

\bibitem[{\citenamefont{{Robinson} and {Baker}}(2000)}]{Robinson_Baker}
\bibinfo{author}{\bibfnamefont{J.}~\bibnamefont{{Robinson}}} \bibnamefont{and}
  \bibinfo{author}{\bibfnamefont{J.~E.} \bibnamefont{{Baker}}},
  \bibinfo{journal}{\mnras} \textbf{\bibinfo{volume}{311}},
  \bibinfo{pages}{781} (\bibinfo{year}{2000}), \eprint{arXiv:astro-ph/9905098}.

\bibitem[{\citenamefont{{Robinson} et~al.}(2000)\citenamefont{{Robinson},
  {Gawiser}, and {Silk}}}]{Robinson_Gawiser_Silk}
\bibinfo{author}{\bibfnamefont{J.}~\bibnamefont{{Robinson}}},
  \bibinfo{author}{\bibfnamefont{E.}~\bibnamefont{{Gawiser}}},
  \bibnamefont{and} \bibinfo{author}{\bibfnamefont{J.}~\bibnamefont{{Silk}}},
  \bibinfo{journal}{\apj} \textbf{\bibinfo{volume}{532}}, \bibinfo{pages}{1}
  (\bibinfo{year}{2000}), \eprint{arXiv:astro-ph/9906156}.

\bibitem[{\citenamefont{{Matarrese} et~al.}(2000)\citenamefont{{Matarrese},
  {Verde}, and {Jimenez}}}]{MVJ}
\bibinfo{author}{\bibfnamefont{S.}~\bibnamefont{{Matarrese}}},
  \bibinfo{author}{\bibfnamefont{L.}~\bibnamefont{{Verde}}}, \bibnamefont{and}
  \bibinfo{author}{\bibfnamefont{R.}~\bibnamefont{{Jimenez}}},
  \bibinfo{journal}{\apj} \textbf{\bibinfo{volume}{541}}, \bibinfo{pages}{10}
  (\bibinfo{year}{2000}), \eprint{arXiv:astro-ph/0001366}.

\bibitem[{\citenamefont{{Press} and {Schechter}}(1974)}]{press-schechter}
\bibinfo{author}{\bibfnamefont{W.~H.} \bibnamefont{{Press}}} \bibnamefont{and}
  \bibinfo{author}{\bibfnamefont{P.}~\bibnamefont{{Schechter}}},
  \bibinfo{journal}{\apj} \textbf{\bibinfo{volume}{187}}, \bibinfo{pages}{425}
  (\bibinfo{year}{1974}).

\bibitem[{\citenamefont{{Kang} et~al.}(2007)\citenamefont{{Kang}, {Norberg},
  and {Silk}}}]{Kang}
\bibinfo{author}{\bibfnamefont{X.}~\bibnamefont{{Kang}}},
  \bibinfo{author}{\bibfnamefont{P.}~\bibnamefont{{Norberg}}},
  \bibnamefont{and} \bibinfo{author}{\bibfnamefont{J.}~\bibnamefont{{Silk}}},
  \bibinfo{journal}{\mnras} \textbf{\bibinfo{volume}{376}},
  \bibinfo{pages}{343} (\bibinfo{year}{2007}), \eprint{arXiv:astro-ph/0701131}.

\bibitem[{\citenamefont{{Grossi} et~al.}(2007)\citenamefont{{Grossi}, {Dolag},
  {Branchini}, {Matarrese}, and {Moscardini}}}]{Grossi}
\bibinfo{author}{\bibfnamefont{M.}~\bibnamefont{{Grossi}}},
  \bibinfo{author}{\bibfnamefont{K.}~\bibnamefont{{Dolag}}},
  \bibinfo{author}{\bibfnamefont{E.}~\bibnamefont{{Branchini}}},
  \bibinfo{author}{\bibfnamefont{S.}~\bibnamefont{{Matarrese}}},
  \bibnamefont{and}
  \bibinfo{author}{\bibfnamefont{L.}~\bibnamefont{{Moscardini}}},
  \bibinfo{journal}{ArXiv e-prints} \textbf{\bibinfo{volume}{707}}
  (\bibinfo{year}{2007}), \eprint{0707.2516}.

\bibitem[{\citenamefont{{Komatsu} and {Spergel}}(2001)}]{komatsu}
\bibinfo{author}{\bibfnamefont{E.}~\bibnamefont{{Komatsu}}} \bibnamefont{and}
  \bibinfo{author}{\bibfnamefont{D.~N.} \bibnamefont{{Spergel}}},
  \bibinfo{journal}{\prd} \textbf{\bibinfo{volume}{63}},
  \bibinfo{pages}{063002} (\bibinfo{year}{2001}),
  \eprint{arXiv:astro-ph/0005036}.

\bibitem[{\citenamefont{{Shirokov} and {Bertschinger}}(2005)}]{gracos}
\bibinfo{author}{\bibfnamefont{A.}~\bibnamefont{{Shirokov}}} \bibnamefont{and}
  \bibinfo{author}{\bibfnamefont{E.}~\bibnamefont{{Bertschinger}}},
  \bibinfo{journal}{ArXiv Astrophysics e-prints}  (\bibinfo{year}{2005}),
  \eprint{astro-ph/0505087}.

\bibitem[{\citenamefont{Shirokov}(2005)}]{alexthesis}
\bibinfo{author}{\bibfnamefont{A.~V.} \bibnamefont{Shirokov}}, Ph.D. thesis,
  \bibinfo{school}{Massachusetts Institute of Technology},
  \bibinfo{address}{Cambridge MA} (\bibinfo{year}{2005}).

\bibitem[{\citenamefont{{Padmanabhan}}(1993)}]{padmanabhan}
\bibinfo{author}{\bibfnamefont{T.}~\bibnamefont{{Padmanabhan}}},
  \emph{\bibinfo{title}{{Structure formation in the universe}}}
  (\bibinfo{publisher}{Cambridge ; New York : Cambridge University Press,
  1993.}, \bibinfo{year}{1993}).

\bibitem[{\citenamefont{{Warren} et~al.}(2006)\citenamefont{{Warren},
  {Abazajian}, {Holz}, and {Teodoro}}}]{warren}
\bibinfo{author}{\bibfnamefont{M.~S.} \bibnamefont{{Warren}}},
  \bibinfo{author}{\bibfnamefont{K.}~\bibnamefont{{Abazajian}}},
  \bibinfo{author}{\bibfnamefont{D.~E.} \bibnamefont{{Holz}}},
  \bibnamefont{and}
  \bibinfo{author}{\bibfnamefont{L.}~\bibnamefont{{Teodoro}}},
  \bibinfo{journal}{\apj} \textbf{\bibinfo{volume}{646}}, \bibinfo{pages}{881}
  (\bibinfo{year}{2006}), \eprint{arXiv:astro-ph/0506395}.

\bibitem[{\citenamefont{{Davis} et~al.}(1985)\citenamefont{{Davis},
  {Efstathiou}, {Frenk}, and {White}}}]{davis}
\bibinfo{author}{\bibfnamefont{M.}~\bibnamefont{{Davis}}},
  \bibinfo{author}{\bibfnamefont{G.}~\bibnamefont{{Efstathiou}}},
  \bibinfo{author}{\bibfnamefont{C.~S.} \bibnamefont{{Frenk}}},
  \bibnamefont{and} \bibinfo{author}{\bibfnamefont{S.~D.~M.}
  \bibnamefont{{White}}}, \bibinfo{journal}{\apj}
  \textbf{\bibinfo{volume}{292}}, \bibinfo{pages}{371} (\bibinfo{year}{1985}).

\bibitem[{\citenamefont{{Jenkins} et~al.}(2001)\citenamefont{{Jenkins},
  {Frenk}, {White}, {Colberg}, {Cole}, {Evrard}, {Couchman}, and
  {Yoshida}}}]{jenkins}
\bibinfo{author}{\bibfnamefont{A.}~\bibnamefont{{Jenkins}}},
  \bibinfo{author}{\bibfnamefont{C.~S.} \bibnamefont{{Frenk}}},
  \bibinfo{author}{\bibfnamefont{S.~D.~M.} \bibnamefont{{White}}},
  \bibinfo{author}{\bibfnamefont{J.~M.} \bibnamefont{{Colberg}}},
  \bibinfo{author}{\bibfnamefont{S.}~\bibnamefont{{Cole}}},
  \bibinfo{author}{\bibfnamefont{A.~E.} \bibnamefont{{Evrard}}},
  \bibinfo{author}{\bibfnamefont{H.~M.~P.} \bibnamefont{{Couchman}}},
  \bibnamefont{and}
  \bibinfo{author}{\bibfnamefont{N.}~\bibnamefont{{Yoshida}}},
  \bibinfo{journal}{\mnras} \textbf{\bibinfo{volume}{321}},
  \bibinfo{pages}{372} (\bibinfo{year}{2001}), \eprint{arXiv:astro-ph/0005260}.

\bibitem[{\citenamefont{{Bond} et~al.}(1991)\citenamefont{{Bond}, {Cole},
  {Efstathiou}, and {Kaiser}}}]{bcek}
\bibinfo{author}{\bibfnamefont{J.~R.} \bibnamefont{{Bond}}},
  \bibinfo{author}{\bibfnamefont{S.}~\bibnamefont{{Cole}}},
  \bibinfo{author}{\bibfnamefont{G.}~\bibnamefont{{Efstathiou}}},
  \bibnamefont{and} \bibinfo{author}{\bibfnamefont{N.}~\bibnamefont{{Kaiser}}},
  \bibinfo{journal}{\apj} \textbf{\bibinfo{volume}{379}}, \bibinfo{pages}{440}
  (\bibinfo{year}{1991}).

\bibitem[{\citenamefont{{Bond} and {Myers}}(1996)}]{bm96}
\bibinfo{author}{\bibfnamefont{J.~R.} \bibnamefont{{Bond}}} \bibnamefont{and}
  \bibinfo{author}{\bibfnamefont{S.~T.} \bibnamefont{{Myers}}},
  \bibinfo{journal}{\apjs} \textbf{\bibinfo{volume}{103}}, \bibinfo{pages}{1}
  (\bibinfo{year}{1996}).

\bibitem[{\citenamefont{{Lukic} et~al.}(2007)\citenamefont{{Lukic}, {Heitmann},
  {Habib}, {Bashinsky}, and {Ricker}}}]{lukic}
\bibinfo{author}{\bibfnamefont{Z.}~\bibnamefont{{Lukic}}},
  \bibinfo{author}{\bibfnamefont{K.}~\bibnamefont{{Heitmann}}},
  \bibinfo{author}{\bibfnamefont{S.}~\bibnamefont{{Habib}}},
  \bibinfo{author}{\bibfnamefont{S.}~\bibnamefont{{Bashinsky}}},
  \bibnamefont{and} \bibinfo{author}{\bibfnamefont{P.~M.}
  \bibnamefont{{Ricker}}}, \bibinfo{journal}{ArXiv Astrophysics e-prints}
  (\bibinfo{year}{2007}), \eprint{astro-ph/0702360}.

\bibitem[{\citenamefont{{Sefusatti} et~al.}(2007)\citenamefont{{Sefusatti},
  {Vale}, {Kadota}, and {Frieman}}}]{Sefusatti}
\bibinfo{author}{\bibfnamefont{E.}~\bibnamefont{{Sefusatti}}},
  \bibinfo{author}{\bibfnamefont{C.}~\bibnamefont{{Vale}}},
  \bibinfo{author}{\bibfnamefont{K.}~\bibnamefont{{Kadota}}}, \bibnamefont{and}
  \bibinfo{author}{\bibfnamefont{J.}~\bibnamefont{{Frieman}}},
  \bibinfo{journal}{\apj} \textbf{\bibinfo{volume}{658}}, \bibinfo{pages}{669}
  (\bibinfo{year}{2007}), \eprint{arXiv:astro-ph/0609124}.

\bibitem[{\citenamefont{{Lacey} and {Cole}}(1993)}]{laceycole}
\bibinfo{author}{\bibfnamefont{C.}~\bibnamefont{{Lacey}}} \bibnamefont{and}
  \bibinfo{author}{\bibfnamefont{S.}~\bibnamefont{{Cole}}},
  \bibinfo{journal}{\mnras} \textbf{\bibinfo{volume}{262}},
  \bibinfo{pages}{627} (\bibinfo{year}{1993}).

\bibitem[{\citenamefont{{Bower}}(1991)}]{bower}
\bibinfo{author}{\bibfnamefont{R.~G.} \bibnamefont{{Bower}}},
  \bibinfo{journal}{\mnras} \textbf{\bibinfo{volume}{248}},
  \bibinfo{pages}{332} (\bibinfo{year}{1991}).

\bibitem[{\citenamefont{{Coles}}(1993)}]{coles93}
\bibinfo{author}{\bibfnamefont{P.}~\bibnamefont{{Coles}}},
  \bibinfo{journal}{\mnras} \textbf{\bibinfo{volume}{262}},
  \bibinfo{pages}{1065} (\bibinfo{year}{1993}).

\bibitem[{\citenamefont{{Fry} and {Gaztanaga}}(1993)}]{fry93}
\bibinfo{author}{\bibfnamefont{J.~N.} \bibnamefont{{Fry}}} \bibnamefont{and}
  \bibinfo{author}{\bibfnamefont{E.}~\bibnamefont{{Gaztanaga}}},
  \bibinfo{journal}{\apj} \textbf{\bibinfo{volume}{413}}, \bibinfo{pages}{447}
  (\bibinfo{year}{1993}), \eprint{arXiv:astro-ph/9302009}.

\bibitem[{\citenamefont{{Scherrer} and {Weinberg}}(1998)}]{scherrer98}
\bibinfo{author}{\bibfnamefont{R.~J.} \bibnamefont{{Scherrer}}}
  \bibnamefont{and} \bibinfo{author}{\bibfnamefont{D.~H.}
  \bibnamefont{{Weinberg}}}, \bibinfo{journal}{\apj}
  \textbf{\bibinfo{volume}{504}}, \bibinfo{pages}{607} (\bibinfo{year}{1998}),
  \eprint{arXiv:astro-ph/9712192}.

\bibitem[{\citenamefont{{Smith}
  et~al.}(2007{\natexlab{a}})\citenamefont{{Smith}, {Scoccimarro}, and
  {Sheth}}}]{smith07a}
\bibinfo{author}{\bibfnamefont{R.~E.} \bibnamefont{{Smith}}},
  \bibinfo{author}{\bibfnamefont{R.}~\bibnamefont{{Scoccimarro}}},
  \bibnamefont{and} \bibinfo{author}{\bibfnamefont{R.~K.}
  \bibnamefont{{Sheth}}}, \bibinfo{journal}{\prd}
  \textbf{\bibinfo{volume}{75}}, \bibinfo{pages}{063512}
  (\bibinfo{year}{2007}{\natexlab{a}}), \eprint{arXiv:astro-ph/0609547}.

\bibitem[{\citenamefont{{Gunn} and {Gott}}(1972)}]{GunnGott72}
\bibinfo{author}{\bibfnamefont{J.~E.} \bibnamefont{{Gunn}}} \bibnamefont{and}
  \bibinfo{author}{\bibfnamefont{J.~R.~I.} \bibnamefont{{Gott}}},
  \bibinfo{journal}{\apj} \textbf{\bibinfo{volume}{176}}, \bibinfo{pages}{1}
  (\bibinfo{year}{1972}).

\bibitem[{\citenamefont{{Blake} et~al.}(2006)\citenamefont{{Blake},
  {Parkinson}, {Bassett}, {Glazebrook}, {Kunz}, and {Nichol}}}]{blake06}
\bibinfo{author}{\bibfnamefont{C.}~\bibnamefont{{Blake}}},
  \bibinfo{author}{\bibfnamefont{D.}~\bibnamefont{{Parkinson}}},
  \bibinfo{author}{\bibfnamefont{B.}~\bibnamefont{{Bassett}}},
  \bibinfo{author}{\bibfnamefont{K.}~\bibnamefont{{Glazebrook}}},
  \bibinfo{author}{\bibfnamefont{M.}~\bibnamefont{{Kunz}}}, \bibnamefont{and}
  \bibinfo{author}{\bibfnamefont{R.~C.} \bibnamefont{{Nichol}}},
  \bibinfo{journal}{\mnras} \textbf{\bibinfo{volume}{365}},
  \bibinfo{pages}{255} (\bibinfo{year}{2006}), \eprint{arXiv:astro-ph/0510239}.

\bibitem[{\citenamefont{{Padmanabhan} et~al.}(2007)\citenamefont{{Padmanabhan},
  {Schlegel}, {Seljak}, {Makarov}, {Bahcall}, {Blanton}, {Brinkmann},
  {Eisenstein}, {Finkbeiner}, {Gunn} et~al.}}]{LRG07}
\bibinfo{author}{\bibfnamefont{N.}~\bibnamefont{{Padmanabhan}}},
  \bibinfo{author}{\bibfnamefont{D.~J.} \bibnamefont{{Schlegel}}},
  \bibinfo{author}{\bibfnamefont{U.}~\bibnamefont{{Seljak}}},
  \bibinfo{author}{\bibfnamefont{A.}~\bibnamefont{{Makarov}}},
  \bibinfo{author}{\bibfnamefont{N.~A.} \bibnamefont{{Bahcall}}},
  \bibinfo{author}{\bibfnamefont{M.~R.} \bibnamefont{{Blanton}}},
  \bibinfo{author}{\bibfnamefont{J.}~\bibnamefont{{Brinkmann}}},
  \bibinfo{author}{\bibfnamefont{D.~J.} \bibnamefont{{Eisenstein}}},
  \bibinfo{author}{\bibfnamefont{D.~P.} \bibnamefont{{Finkbeiner}}},
  \bibinfo{author}{\bibfnamefont{J.~E.} \bibnamefont{{Gunn}}},
  \bibnamefont{et~al.}, \bibinfo{journal}{\mnras}
  \textbf{\bibinfo{volume}{378}}, \bibinfo{pages}{852} (\bibinfo{year}{2007}),
  \eprint{arXiv:astro-ph/0605302}.

\bibitem[{\citenamefont{{Eisenstein} and {Hu}}(1999)}]{Hu_transfer}
\bibinfo{author}{\bibfnamefont{D.~J.} \bibnamefont{{Eisenstein}}}
  \bibnamefont{and} \bibinfo{author}{\bibfnamefont{W.}~\bibnamefont{{Hu}}},
  \bibinfo{journal}{\apj} \textbf{\bibinfo{volume}{511}}, \bibinfo{pages}{5}
  (\bibinfo{year}{1999}), \eprint{arXiv:astro-ph/9710252}.

\bibitem[{\citenamefont{Crocce and Scoccimarro}(2007)}]{Crocce:2007dt}
\bibinfo{author}{\bibfnamefont{M.}~\bibnamefont{Crocce}} \bibnamefont{and}
  \bibinfo{author}{\bibfnamefont{R.}~\bibnamefont{Scoccimarro}}
  (\bibinfo{year}{2007}), \eprint{arXiv:0704.2783 [astro-ph]}.

\bibitem[{\citenamefont{{Smith}
  et~al.}(2007{\natexlab{b}})\citenamefont{{Smith}, {Scoccimarro}, and
  {Sheth}}}]{smith07b}
\bibinfo{author}{\bibfnamefont{R.~E.} \bibnamefont{{Smith}}},
  \bibinfo{author}{\bibfnamefont{R.}~\bibnamefont{{Scoccimarro}}},
  \bibnamefont{and} \bibinfo{author}{\bibfnamefont{R.~K.}
  \bibnamefont{{Sheth}}}, \bibinfo{journal}{ArXiv Astrophysics e-prints}
  (\bibinfo{year}{2007}{\natexlab{b}}), \eprint{astro-ph/0703620}.

\bibitem[{\citenamefont{Sachs and Wolfe}(1967)}]{Sachs:1967er}
\bibinfo{author}{\bibfnamefont{R.~K.} \bibnamefont{Sachs}} \bibnamefont{and}
  \bibinfo{author}{\bibfnamefont{A.~M.} \bibnamefont{Wolfe}},
  \bibinfo{journal}{Astrophys. J.} \textbf{\bibinfo{volume}{147}},
  \bibinfo{pages}{73} (\bibinfo{year}{1967}).

\bibitem[{\citenamefont{Crittenden and Turok}(1996)}]{Crittenden:1995ak}
\bibinfo{author}{\bibfnamefont{R.~G.} \bibnamefont{Crittenden}}
  \bibnamefont{and} \bibinfo{author}{\bibfnamefont{N.}~\bibnamefont{Turok}},
  \bibinfo{journal}{Phys. Rev. Lett.} \textbf{\bibinfo{volume}{76}},
  \bibinfo{pages}{575} (\bibinfo{year}{1996}), \eprint{astro-ph/9510072}.

\bibitem[{\citenamefont{Bean and Dore}(2004)}]{Bean:2003fb}
\bibinfo{author}{\bibfnamefont{R.}~\bibnamefont{Bean}} \bibnamefont{and}
  \bibinfo{author}{\bibfnamefont{O.}~\bibnamefont{Dore}},
  \bibinfo{journal}{Phys. Rev.} \textbf{\bibinfo{volume}{D69}},
  \bibinfo{pages}{083503} (\bibinfo{year}{2004}), \eprint{astro-ph/0307100}.

\bibitem[{\citenamefont{{Boughn} and {Crittenden}}(2004)}]{boughn04}
\bibinfo{author}{\bibfnamefont{S.}~\bibnamefont{{Boughn}}} \bibnamefont{and}
  \bibinfo{author}{\bibfnamefont{R.}~\bibnamefont{{Crittenden}}},
  \bibinfo{journal}{\nat} \textbf{\bibinfo{volume}{427}}, \bibinfo{pages}{45}
  (\bibinfo{year}{2004}).

\bibitem[{\citenamefont{{Nolta}~\etal}(2004)}]{nolta04}
\bibinfo{author}{\bibfnamefont{M.~R.} \bibnamefont{{Nolta}~\etal}},
  \bibinfo{journal}{\apj} \textbf{\bibinfo{volume}{608}}, \bibinfo{pages}{10}
  (\bibinfo{year}{2004}), \eprint{astro-ph/0305097}.

\bibitem[{\citenamefont{{Fosalba}~\etal}(2003)}]{fosalba03}
\bibinfo{author}{\bibfnamefont{P.}~\bibnamefont{{Fosalba}~\etal}},
  \bibinfo{journal}{\apjl} \textbf{\bibinfo{volume}{597}}, \bibinfo{pages}{L89}
  (\bibinfo{year}{2003}).

\bibitem[{\citenamefont{{Scranton}~\etal}(2003)}]{scranton03}
\bibinfo{author}{\bibfnamefont{R.}~\bibnamefont{{Scranton}~\etal}},
  \bibinfo{journal}{ArXiv Astrophysics e-prints}  (\bibinfo{year}{2003}),
  \eprint{astro-ph/0307335}.

\bibitem[{\citenamefont{{Fosalba} and {Gazta{\~n}aga}}(2004)}]{fosalba04}
\bibinfo{author}{\bibfnamefont{P.}~\bibnamefont{{Fosalba}}} \bibnamefont{and}
  \bibinfo{author}{\bibfnamefont{E.}~\bibnamefont{{Gazta{\~n}aga}}},
  \bibinfo{journal}{\mnras} \textbf{\bibinfo{volume}{350}},
  \bibinfo{pages}{L37} (\bibinfo{year}{2004}), \eprint{astro-ph/0305468}.

\bibitem[{\citenamefont{Padmanabhan
  et~al.}(2005{\natexlab{a}})}]{Padmanabhan:2004fy}
\bibinfo{author}{\bibfnamefont{N.}~\bibnamefont{Padmanabhan}}
  \bibnamefont{et~al.}, \bibinfo{journal}{Phys. Rev.}
  \textbf{\bibinfo{volume}{D72}}, \bibinfo{pages}{043525}
  (\bibinfo{year}{2005}{\natexlab{a}}), \eprint{astro-ph/0410360}.

\bibitem[{\citenamefont{{Afshordi} et~al.}(2004)\citenamefont{{Afshordi},
  {Loh}, and {Strauss}}}]{afshordi04a}
\bibinfo{author}{\bibfnamefont{N.}~\bibnamefont{{Afshordi}}},
  \bibinfo{author}{\bibfnamefont{Y.-S.} \bibnamefont{{Loh}}}, \bibnamefont{and}
  \bibinfo{author}{\bibfnamefont{M.~A.} \bibnamefont{{Strauss}}},
  \bibinfo{journal}{\prd} \textbf{\bibinfo{volume}{69}},
  \bibinfo{pages}{083524} (\bibinfo{year}{2004}), \eprint{astro-ph/0308260}.

\bibitem[{\citenamefont{{Cabre}~\etal}(2006)}]{cabre06}
\bibinfo{author}{\bibfnamefont{A.}~\bibnamefont{{Cabre}~\etal}},
  \bibinfo{journal}{ArXiv Astrophysics e-prints}  (\bibinfo{year}{2006}),
  \eprint{astro-ph/0603690}.

\bibitem[{\citenamefont{{Scranton}~\etal}(2007)}]{scranton07}
\bibinfo{author}{\bibfnamefont{R.}~\bibnamefont{{Scranton}~\etal}},
  \bibinfo{journal}{in preparation}  (\bibinfo{year}{2007}).

\bibitem[{\citenamefont{Afshordi}(2004)}]{Afshordi:2004kz}
\bibinfo{author}{\bibfnamefont{N.}~\bibnamefont{Afshordi}},
  \bibinfo{journal}{Phys. Rev.} \textbf{\bibinfo{volume}{D70}},
  \bibinfo{pages}{083536} (\bibinfo{year}{2004}), \eprint{astro-ph/0401166}.

\bibitem[{\citenamefont{Hu and Scranton}(2004)}]{Hu:2004yd}
\bibinfo{author}{\bibfnamefont{W.}~\bibnamefont{Hu}} \bibnamefont{and}
  \bibinfo{author}{\bibfnamefont{R.}~\bibnamefont{Scranton}},
  \bibinfo{journal}{Phys. Rev.} \textbf{\bibinfo{volume}{D70}},
  \bibinfo{pages}{123002} (\bibinfo{year}{2004}), \eprint{astro-ph/0408456}.

\bibitem[{\citenamefont{Smith and
  Zaldarriaga}(2006{\natexlab{b}})}]{Smith:2006ud}
\bibinfo{author}{\bibfnamefont{K.~M.} \bibnamefont{Smith}} \bibnamefont{and}
  \bibinfo{author}{\bibfnamefont{M.}~\bibnamefont{Zaldarriaga}}
  (\bibinfo{year}{2006}{\natexlab{b}}), \eprint{astro-ph/0612571}.

\bibitem[{\citenamefont{{Sadeh} et~al.}(2007)\citenamefont{{Sadeh}, {Rephaeli},
  and {Silk}}}]{sadeh07}
\bibinfo{author}{\bibfnamefont{S.}~\bibnamefont{{Sadeh}}},
  \bibinfo{author}{\bibfnamefont{Y.}~\bibnamefont{{Rephaeli}}},
  \bibnamefont{and} \bibinfo{author}{\bibfnamefont{J.}~\bibnamefont{{Silk}}},
  \bibinfo{journal}{\mnras} \textbf{\bibinfo{volume}{380}},
  \bibinfo{pages}{637} (\bibinfo{year}{2007}), \eprint{arXiv:0706.1340}.

\bibitem[{\citenamefont{Cooray et~al.}(2004)\citenamefont{Cooray, Huterer, and
  Baumann}}]{Coo_Hut_Bau}
\bibinfo{author}{\bibfnamefont{A.}~\bibnamefont{Cooray}},
  \bibinfo{author}{\bibfnamefont{D.}~\bibnamefont{Huterer}}, \bibnamefont{and}
  \bibinfo{author}{\bibfnamefont{D.}~\bibnamefont{Baumann}},
  \bibinfo{journal}{Phys. Rev.} \textbf{\bibinfo{volume}{D69}},
  \bibinfo{pages}{027301} (\bibinfo{year}{2004}), \eprint{astro-ph/0304268}.

\bibitem[{\citenamefont{Holder et~al.}(2001)\citenamefont{Holder, Haiman, and
  Mohr}}]{Holder_Haiman_Mohr}
\bibinfo{author}{\bibfnamefont{G.}~\bibnamefont{Holder}},
  \bibinfo{author}{\bibfnamefont{Z.}~\bibnamefont{Haiman}}, \bibnamefont{and}
  \bibinfo{author}{\bibfnamefont{J.}~\bibnamefont{Mohr}},
  \bibinfo{journal}{Astrophys. J.} \textbf{\bibinfo{volume}{560}},
  \bibinfo{pages}{L111} (\bibinfo{year}{2001}), \eprint{astro-ph/0105396}.

\bibitem[{\citenamefont{Huterer et~al.}(2004)\citenamefont{Huterer, Kim,
  Krauss, and Broderick}}]{Huterer_Broderick}
\bibinfo{author}{\bibfnamefont{D.}~\bibnamefont{Huterer}},
  \bibinfo{author}{\bibfnamefont{A.}~\bibnamefont{Kim}},
  \bibinfo{author}{\bibfnamefont{L.~M.} \bibnamefont{Krauss}},
  \bibnamefont{and}
  \bibinfo{author}{\bibfnamefont{T.}~\bibnamefont{Broderick}},
  \bibinfo{journal}{Astrophys. J.} \textbf{\bibinfo{volume}{615}},
  \bibinfo{pages}{595} (\bibinfo{year}{2004}), \eprint{astro-ph/0402002}.

\bibitem[{\citenamefont{Lima and Hu}(2004)}]{Lima_Hu_04}
\bibinfo{author}{\bibfnamefont{M.}~\bibnamefont{Lima}} \bibnamefont{and}
  \bibinfo{author}{\bibfnamefont{W.}~\bibnamefont{Hu}}, \bibinfo{journal}{Phys.
  Rev.} \textbf{\bibinfo{volume}{D70}}, \bibinfo{pages}{043504}
  (\bibinfo{year}{2004}), \eprint{astro-ph/0401559}.

\bibitem[{\citenamefont{Padmanabhan
  et~al.}(2005{\natexlab{b}})}]{Padmanabhan:2004ic}
\bibinfo{author}{\bibfnamefont{N.}~\bibnamefont{Padmanabhan}}
  \bibnamefont{et~al.} (\bibinfo{collaboration}{SDSS}), \bibinfo{journal}{Mon.
  Not. Roy. Astron. Soc.} \textbf{\bibinfo{volume}{359}}, \bibinfo{pages}{237}
  (\bibinfo{year}{2005}{\natexlab{b}}), \eprint{astro-ph/0407594}.

\bibitem[{\citenamefont{Ilbert et~al.}(2006)}]{Ilbert:2006dp}
\bibinfo{author}{\bibfnamefont{O.}~\bibnamefont{Ilbert}} \bibnamefont{et~al.}
  (\bibinfo{year}{2006}), \eprint{astro-ph/0603217}.

\bibitem[{\citenamefont{{Avila-Reese} et~al.}(2003)\citenamefont{{Avila-Reese},
  {Col{\'{\i}}n}, {Piccinelli}, and {Firmani}}}]{avila03}
\bibinfo{author}{\bibfnamefont{V.}~\bibnamefont{{Avila-Reese}}},
  \bibinfo{author}{\bibfnamefont{P.}~\bibnamefont{{Col{\'{\i}}n}}},
  \bibinfo{author}{\bibfnamefont{G.}~\bibnamefont{{Piccinelli}}},
  \bibnamefont{and}
  \bibinfo{author}{\bibfnamefont{C.}~\bibnamefont{{Firmani}}},
  \bibinfo{journal}{\apj} \textbf{\bibinfo{volume}{598}}, \bibinfo{pages}{36}
  (\bibinfo{year}{2003}), \eprint{arXiv:astro-ph/0306293}.

\bibitem[{\citenamefont{{Gladders} et~al.}(2003)\citenamefont{{Gladders},
  {Hoekstra}, {Yee}, {Hall}, and {Barrientos}}}]{rcs}
\bibinfo{author}{\bibfnamefont{M.~D.} \bibnamefont{{Gladders}}},
  \bibinfo{author}{\bibfnamefont{H.}~\bibnamefont{{Hoekstra}}},
  \bibinfo{author}{\bibfnamefont{H.~K.~C.} \bibnamefont{{Yee}}},
  \bibinfo{author}{\bibfnamefont{P.~B.} \bibnamefont{{Hall}}},
  \bibnamefont{and} \bibinfo{author}{\bibfnamefont{L.~F.}
  \bibnamefont{{Barrientos}}}, \bibinfo{journal}{\apj}
  \textbf{\bibinfo{volume}{593}}, \bibinfo{pages}{48} (\bibinfo{year}{2003}),
  \eprint{arXiv:astro-ph/0303341}.

\bibitem[{\citenamefont{{Dalal} et~al.}(2004)\citenamefont{{Dalal}, {Holder},
  and {Hennawi}}}]{arcs04}
\bibinfo{author}{\bibfnamefont{N.}~\bibnamefont{{Dalal}}},
  \bibinfo{author}{\bibfnamefont{G.}~\bibnamefont{{Holder}}}, \bibnamefont{and}
  \bibinfo{author}{\bibfnamefont{J.~F.} \bibnamefont{{Hennawi}}},
  \bibinfo{journal}{\apj} \textbf{\bibinfo{volume}{609}}, \bibinfo{pages}{50}
  (\bibinfo{year}{2004}), \eprint{arXiv:astro-ph/0310306}.

\bibitem[{\citenamefont{{Hennawi} et~al.}(2007)\citenamefont{{Hennawi},
  {Dalal}, {Bode}, and {Ostriker}}}]{hennawi07}
\bibinfo{author}{\bibfnamefont{J.~F.} \bibnamefont{{Hennawi}}},
  \bibinfo{author}{\bibfnamefont{N.}~\bibnamefont{{Dalal}}},
  \bibinfo{author}{\bibfnamefont{P.}~\bibnamefont{{Bode}}}, \bibnamefont{and}
  \bibinfo{author}{\bibfnamefont{J.~P.} \bibnamefont{{Ostriker}}},
  \bibinfo{journal}{\apj} \textbf{\bibinfo{volume}{654}}, \bibinfo{pages}{714}
  (\bibinfo{year}{2007}), \eprint{arXiv:astro-ph/0506171}.

\bibitem[{\citenamefont{{Broadhurst} and {Barkana}}(2008)}]{broadhurst08}
\bibinfo{author}{\bibfnamefont{T.}~\bibnamefont{{Broadhurst}}}
  \bibnamefont{and}
  \bibinfo{author}{\bibfnamefont{R.}~\bibnamefont{{Barkana}}},
  \bibinfo{journal}{ArXiv e-prints} \textbf{\bibinfo{volume}{801}}
  (\bibinfo{year}{2008}), \eprint{0801.1875}.

\bibitem[{\citenamefont{{McDonald} et~al.}(2006)\citenamefont{{McDonald},
  {Seljak}, {Burles}, {Schlegel}, {Weinberg}, {Cen}, {Shih}, {Schaye},
  {Schneider}, {Bahcall} et~al.}}]{sdss_lyaf}
\bibinfo{author}{\bibfnamefont{P.}~\bibnamefont{{McDonald}}},
  \bibinfo{author}{\bibfnamefont{U.}~\bibnamefont{{Seljak}}},
  \bibinfo{author}{\bibfnamefont{S.}~\bibnamefont{{Burles}}},
  \bibinfo{author}{\bibfnamefont{D.~J.} \bibnamefont{{Schlegel}}},
  \bibinfo{author}{\bibfnamefont{D.~H.} \bibnamefont{{Weinberg}}},
  \bibinfo{author}{\bibfnamefont{R.}~\bibnamefont{{Cen}}},
  \bibinfo{author}{\bibfnamefont{D.}~\bibnamefont{{Shih}}},
  \bibinfo{author}{\bibfnamefont{J.}~\bibnamefont{{Schaye}}},
  \bibinfo{author}{\bibfnamefont{D.~P.} \bibnamefont{{Schneider}}},
  \bibinfo{author}{\bibfnamefont{N.~A.} \bibnamefont{{Bahcall}}},
  \bibnamefont{et~al.}, \bibinfo{journal}{\apjs}
  \textbf{\bibinfo{volume}{163}}, \bibinfo{pages}{80} (\bibinfo{year}{2006}),
  \eprint{arXiv:astro-ph/0405013}.

\bibitem[{\citenamefont{{Kaiser}}(1984)}]{kaiser84}
\bibinfo{author}{\bibfnamefont{N.}~\bibnamefont{{Kaiser}}},
  \bibinfo{journal}{\apjl} \textbf{\bibinfo{volume}{284}}, \bibinfo{pages}{L9}
  (\bibinfo{year}{1984}).

\bibitem[{\citenamefont{{Bardeen} et~al.}(1986)\citenamefont{{Bardeen}, {Bond},
  {Kaiser}, and {Szalay}}}]{bbks}
\bibinfo{author}{\bibfnamefont{J.~M.} \bibnamefont{{Bardeen}}},
  \bibinfo{author}{\bibfnamefont{J.~R.} \bibnamefont{{Bond}}},
  \bibinfo{author}{\bibfnamefont{N.}~\bibnamefont{{Kaiser}}}, \bibnamefont{and}
  \bibinfo{author}{\bibfnamefont{A.~.~S.} \bibnamefont{{Szalay}}},
  \bibinfo{journal}{\apj} \textbf{\bibinfo{volume}{304}}, \bibinfo{pages}{15}
  (\bibinfo{year}{1986}).

\end{thebibliography}

\end{document}